\documentclass[aps,amsfonts,prb,twocolumn,superscriptaddress,floatfix]{revtex4-1}

\usepackage{epsfig}
\usepackage{amsmath}
\usepackage{amssymb}
\usepackage{bm}
\usepackage{bbm}
\usepackage{graphicx}
\usepackage[usenames,dvipsnames,svgnames,table]{xcolor}
\usepackage{ulem}

\definecolor{darkgreen}{rgb}{0,0.4,0}

\newcommand{\beq}{\begin{equation}}
\newcommand{\eeq}{\end{equation}}
\newcommand{\beqarray}{\begin{eqnarray}}
\newcommand{\eeqarray}{\end{eqnarray}}
\newcommand{\Hc}{\ensuremath{\mbox{H.c.}}} % Hermitian conjugate
\newcommand{\eq}[1]{Eq.~\eqref{#1}} % Eq. label
\newcommand{\fig}[1]{Fig.~\ref{#1}} % Fig. label
\newcommand{\Ref}[1]{Ref.~\onlinecite{#1}} % Ref. label

\newcommand{\ignore}[1]{}

\begin{document}

\allowdisplaybreaks

\title{Loop currents and anomalous Hall effect from 
    time-reversal symmetry-breaking
superconductivity on the honeycomb lattice}
\author{P. M. R. Brydon}
\email{philip.brydon@otago.ac.nz}
\affiliation{Department of Physics and MacDiarmid Institute for Advanced Materials and Nanotechnology, University of Otago, PO Box 56,
  Dunedin 9054, New Zealand}
\author{D. S. L. Abergel}
\affiliation{Nordita, KTH Royal Institute of Technology and Stockholm
  University, Roslagstullsbacken 23, SE-106 91 Stockholm, Sweden}
\author{D. F. Agterberg}
\affiliation{Department of Physics, University of Wisconsin,
                 Milwaukee, WI 53201, USA}
\author{Victor M. Yakovenko}
\email{yakovenk@physics.umd.edu}
\affiliation{Department of Physics, CMTC and JQI, University of
  Maryland, College Park, MD 20742-4111, USA}

\date{\today}

\begin{abstract}

We study a tight-binding model on the honeycomb lattice
of chiral $d$-wave superconductivity  that breaks time-reversal symmetry. 
Due to its nontrivial sublattice structure, we show that it is
possible to construct a gauge-invariant time-reversal-odd 
bilinear of the pairing potential. The existence of this bilinear
reflects the sublattice polarization of the pairing state. We show
that it  generates
persistent loop current correlations around each lattice site
and opens a 
topological mass gap at the Dirac points, resembling Haldane's model
of the anomalous quantum Hall effect. In addition to the usual chiral
$d$-wave edge states, there also exist electron-like edge
resonances due to the topological mass gap. We show that the
  presence of loop-current correlations directly leads to a nonzero
  intrinsic ac Hall conductivity, 
which produces the polar Kerr effect without an external magnetic field.
Similar results also hold for the nearest-neighbor chiral $p$-wave
pairing.  We briefly discuss the relevance of our results to superconductivity in twisted bilayer graphene.
\end{abstract}

\maketitle

%%%%%%%%%%%%%%%%%%%%%%%%%%%%%%%%%%%%%%%%%%%%%%%%
\section{Introduction}
%%%%%%%%%%%%%%%%%%%%%%%%%%%%%%%%%%%%%%%%%%%%%%%%

Chiral superconductors, which possess order parameters that
break time-reversal symmetry, are currently the subject of much attention due
to their nontrivial topological properties.~\cite{SatoRPP2017,KallinRPP2016}
The best known example of a chiral pairing state is the A phase of
superfluid $^3$He.~\cite{LeggettRMP1975} Here Cooper pairs have the
orbital angular momentum quantum numbers $L=1$ and $L_z=\pm 1$, and
the pairing potential has $(p_x\pm ip_y)$-wave symmetry. A direct
solid-state analogue of this phase is believed to be realized in the
triplet superconductor $\rm Sr_2RuO_4$.~\cite{Mackenzie2017} Chiral
superconductivity can also be obtained for pairing with higher-order
orbital angular momentum. For example, the low-temperature 
 superconducting phase of
$\rm UPt_3$ may realize a chiral $f$-wave
state,~\cite{JoyntRMP2002,UPt3chiral} while chiral $d$-wave
superconducting states have been proposed for $\rm URu_2Si_2$,~\cite{Kasahara2007}
$\rm SrPtAs$,~\cite{Fischer2014} and twisted bilayer graphene.~\cite{Cao2018} 
Many other materials have been
predicted to show chiral superconductivity, such as
water-intercalated sodium cobaltate Na$_x$CoO$_2\cdot
y$H$_2$O,~\cite{NaxCoO2papers}  the half-Heusler compound
$\rm YPtBi$,~\cite{Bry2016} and transition metal
dichalcogenides.~\cite{TiSe2,MoS2,dichalcogenides}

The breaking of time-reversal symmetry in a chiral superconductor can be
revealed by a number of experimental techniques, e.g.\ muon spin rotation
or Josephson interferometry.~\cite{KallinRPP2016} In the last dozen
years, measurements of the polar Kerr effect have emerged as a key
experimental probe.~\cite{KapNJP2009} It gives evidence for an
anomalous ac Hall conductivity at zero external magnetic field, which
is a signature of broken time-reversal symmetry. A number of
superconductors have been shown to display a nonzero Kerr signal below
their critical temperatures, specifically
Sr$_2$RuO$_4$,~\cite{Sr2RuO4Kerr}  UPt$_3$,~\cite{UPt3Kerr}
URu$_2$Si$_2$,~\cite{URu2Si2Kerr}
PrOs$_4$Sb$_{12}$,~\cite{PrOs4Sb12Kerr} and Bi/Ni
bilayers.~\cite{BiNiKerr} Although these observations give clear
evidence for broken time-reversal symmetry, there is ongoing debate
over the mechanism underlying the polar Kerr effect in chiral
superconductors. An extrinsic Kerr effect may originate from impurity
scattering,~\cite{Goryo2008,Lutchyn2009,Koenig2017} whereas an
intrinsic Kerr effect is possible for clean multiband  
superconductors.~\cite{Wysokinski2012,Taylor2012,GradhandSr2RuO4,RobbinsSr2RuO4,KomendovaPRL2017,KerrUPt3_theory,JoyntWu2017,Triola2017}  
The latter mechanism requires that the pairing potential depends on
electronic degrees of freedom beyond the usual spin index,
e.g.\  orbital or sublattice indices. However, it remains unclear 
what general model-independent conditions  these additional electronic
degrees of freedom have to satisfy  in order to produce a Kerr effect.  
Here we develop a general condition for this and then apply it to a 
minimal model of a chiral $d$-wave superconductor 
in order to clarify the underlying physics.

Such a minimal theoretical model of a chiral superconductor is provided by
the extended Hubbard model on the honeycomb lattice.~\cite{Baskaran2002,Uchoa2007} 
Various theoretical techniques~\cite{BlaSchDon2007,Honerkamp2008,Pathak2010,Raghu2010,Ma2011,Kiesel2012,Nandkishore2012a,Nandkishore2012b,Wu2013,Gu2013,BlaSchPRB2014,JiangPRX2014,BlaSchHonJPCM2014,XiaoEPL2016,Xu2016,Ying2017}
applied to this system generally agree on the existence of a
spin-singlet chiral $d$-wave state at a doping level close to the van
Hove singularity. Closer to half-filling, however, different methods
have yielded singlet and triplet
pairings,~\cite{Raghu2010,BlaSchPRB2014,Faye2015,XiaoEPL2016,Xu2016,Ying2017,RoyJur2014} pair-density-wave Kekule order,~\cite{Roy2010,Kunst2015} or
an unconventional coexistence with antiferromagnetism.~\cite{BlackSchaeffer2015,Faye2016,QiFuSunGu2017}
The purpose of our paper is not to further interrogate the phase
diagram, but rather to examine the properties of the chiral $d$-wave
state in the case where the nearest-neighbor pairing dominates. Such
inter-sublattice pairing would satisfy the multiband
requirement~\cite{Taylor2012} for the anomalous Hall conductivity. 
Thus, chiral $d$-wave pairing on the honeycomb lattice
provides a minimal model of the intrinsic Kerr effect, in contrast to the more
complicated multiband models of
Sr$_2$RuO$_4$~\cite{Wysokinski2012,Taylor2012,GradhandSr2RuO4,RobbinsSr2RuO4,KomendovaPRL2017} and UPt$_3$.~\cite{KerrUPt3_theory,JoyntWu2017,Triola2017} 
The recent discovery of superconductivity in twisted bilayer
graphene,~\cite{Cao2018} which has been proposed to realize a
chiral $d$-wave state,~\cite{Wu2018,Xu2018,Fidrysiak2018,Guo2018,Su2018,Liu2018}
makes this study timely. We discuss the relationship between our
model and these proposals in more detail near the end of  the
paper.

Using this minimal model as an example, we show how to construct a
gauge-invariant time-reversal-odd term by taking the product of the
pairing potential and its Hermitian conjugate. The existence of such a
bilinear is a prerequisite for the experimental detection of
time-reversal symmetry-breaking superconductivity in a clean and
homogeneous system. In the honeycomb model, the bilinear arises from the
varying participation of the two sublattices in the pairing across the
Brillouin zone and describes spontaneous breaking of the discrete $\mathbb{Z}_2$ time-reversal symmetry.  The presence of this term results in the opening
of a topological mass gap at the Dirac points and the emergence of
persistent loop current correlations, in a striking analogy to
Haldane's model of the anomalous Hall 
insulator.~\cite{Haldane1988} Furthermore, we  show 
that the loop current correlations imply a nonzero anomalous
Hall conductivity, hence 
connecting the polar Kerr effect in superconductors with
the time-reversal-odd bilinear product of the pairing potentials. 

The paper is organized as follows.  We start in Sec.~\ref{sec:model} by introducing the model of spin-singlet chiral $d$-wave pairing on the honeycomb lattice. In
Sec.~\ref{sec:TROB-hex} we define a gauge-invariant bilinear product
of the superconducting pairing potentials that breaks time-reversal
symmetry.  As a consequence of the existence of this bilinear, we
demonstrate the opening of the mass gap at the Dirac point in
Sec.~\ref{sec:massgap} and the existence of loop currents in
Sec.~\ref{sec:loops}.  The anomalous ac Hall conductivity is
calculated in Sec.~\ref{sec:hall}. A phenomenological description of
the loop currents is outlined in Sec.~\ref{sec:phenomeno}.  The
relationship of our work to proposals of chiral $d$-wave
superconductivity in twisted bilayer graphene is discussed in
Sec.~\ref{sec:TBLG}.  We conclude in Sec.~\ref{sec:conclusions} with a
brief discussion of the broader implications of our work. In
Appendix~\ref{sec:chiralp} we present similar results for a
spin-triplet chiral $p$-wave state on the honeycomb lattice.  In
Appendix~\ref{sec:TROBgeneral} we show how the bilinear discussed in
Sec.~\ref{sec:model} applies to a broader class of Hamiltonians.  More
general expressions for the loop-current order and the Hall
conductivity in the case of inequivalent sublattices are given in
Appendix~\ref{sec:genexp}. The high-frequency small-gap limit of the
ac Hall conductivity is derived in Appendix~\ref{eq:diagrams}.

%%%%%%%%%%%%%%%%%%%%%%%%%%%%%%%%%%%%%%%%%%%%%%%%
 \section{\protect Microscopic model}
\label{sec:model}
%%%%%%%%%%%%%%%%%%%%%%%%%%%%%%%%%%%%%%%%%%%%%%%%

The Bogoliubov-de Gennes (BdG) Hamiltonian of 
superconducting pairing on the honeycomb lattice is
\beq
H = \sum_{\bm k}\Psi_{\bm k}^\dagger\left(\begin{array}{cc}
  H_{0}({\bm k}) & \Delta({\bm k}) \\
  \Delta^\dagger({\bm k}) & -H^{T}_{0}(-{\bm
    k})\end{array}\right)\Psi_{\bm k}\,, \label{eq:HBdG}
\eeq
where
$\Psi_{\bm k} = (a^{}_{\bm k,\uparrow},b^{}_{\bm k,\uparrow},
a^\dagger_{-\bm k,\downarrow},b^\dagger_{-\bm k,\downarrow})^T$,  
and the operator $a^{}_{{\bm k},\sigma}$ ($b^{}_{{\bm k},\sigma}$) 
annihilates an electron with momentum ${\bm k}=(k_x,k_y)$ and 
spin $\sigma$ on the $A$ ($B$) sublattices.
In \eq{eq:HBdG}, $H_0(\bm k)$ and $\Delta(\bm k)$ are $2\times2$ matrices 
in the sublattice space, and the absence of spin-orbit coupling allows 
the spin variables to be factored out.

\begin{figure}
\includegraphics[width=0.75\columnwidth]{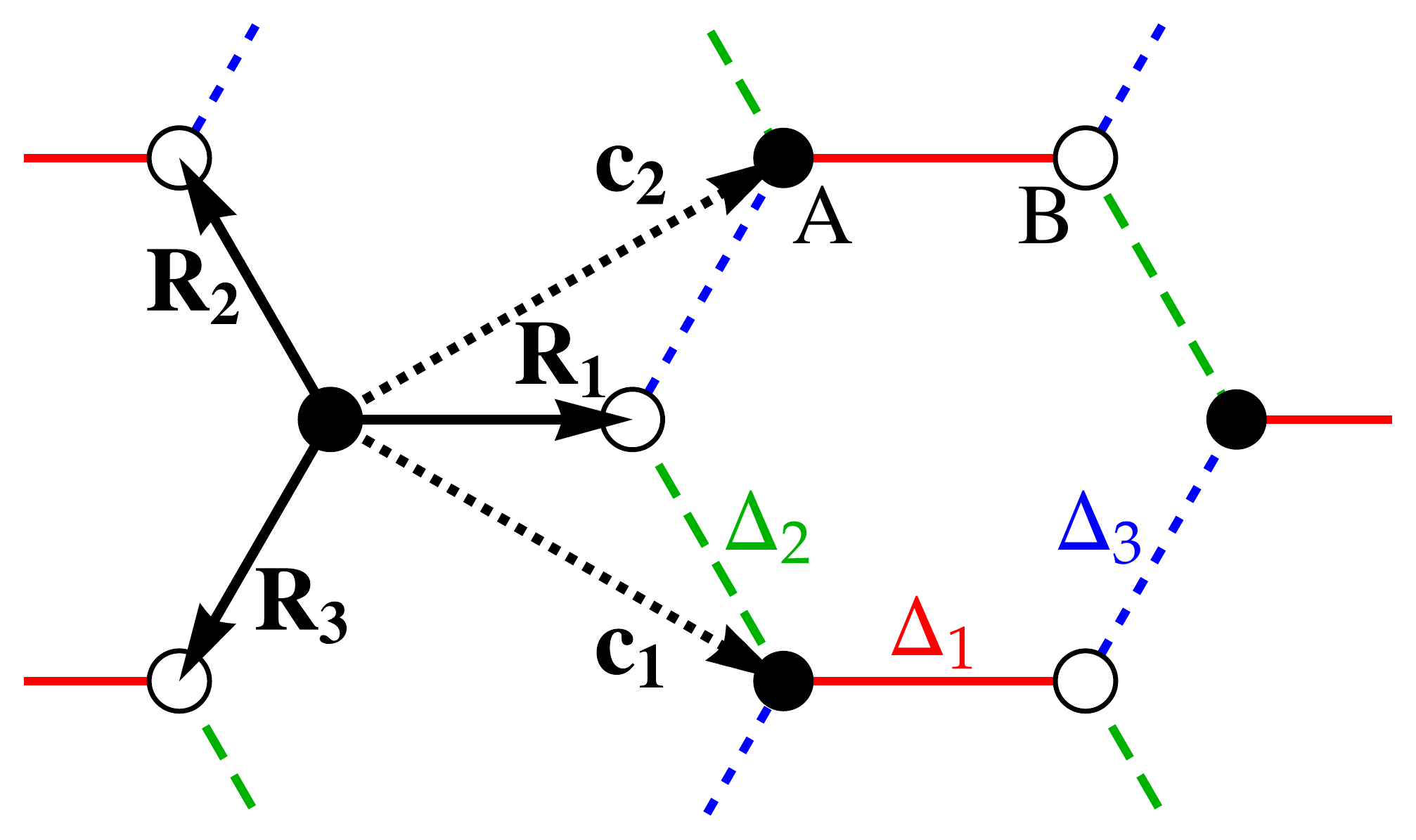}
\caption{(Color online) Schematic diagram of the chiral $d$-wave pairing
  on the honeycomb lattice. The two sublattices are denoted by the black
  ($A$) and white ($B$) circles. The nearest-neighbor vectors 
  ${\bm R}_{j=1,2,3}$ are shown as the black arrows, whereas 
  ${\bm c}_1$ and ${\bm c}_2$ denote the primitive lattice vectors. The
  pairing potentials on the three nearest-neighbor bonds are
  $\{\Delta_1,\Delta_2,\Delta_3\}=
  \Delta\{1,e^{2\pi i/3},e^{4\pi i/3}\}$. The time-reversed pairing
  is obtained by complex conjugation. }
  \label{fig:graphenelattice}
\end{figure}

Using the Pauli matrices $s_\lambda$ to encode the sublattice 
degree of freedom, we write the normal-state Hamiltonian as
\beqarray
&H_{0}({\bm k}) = \epsilon^x_{\bm k}{s}_x + \epsilon^y_{\bm k}{s}_y 
  + \delta_s{s}_z - \mu{s}_0,&  \label{eq:H0k} \\
&\epsilon^x_{\bm k} = -t\sum\limits_{j=1}^3\cos({\bm k}\cdot{\bm R}_j), \;\;
\epsilon^y_{\bm k} = t\sum\limits_{j=1}^3\sin({\bm k}\cdot{\bm R}_j).&  \label{eq:epsilon-xy}
\eeqarray
Here $\mu$ is the chemical potential, $t$ is the nearest-neighbor
hopping  amplitude, and the ${\bm R}_j$ are the vectors of length $a$
connecting an $A$ site to its neighboring $B$ sites, see
\fig{fig:graphenelattice}. For generality, we also include the
Semenoff term~\cite{Semenoff} as a staggered potential
$\delta_s$. This makes the $A$ and $B$ sites inequivalent, hence
breaking inversion symmetry and lowering the point group from $D_{6h}$
to $D_{3h}$. 

%%%%%%%%%%%%%%%%%%%%%%%%%%%%%%%%%%%%%%%%%%%%%%%%
%%%%%%%%%%%%%%%%%%%%%%%%%%%%%%%%%%%%%%%%%%%%%%%%

We consider chiral spin-singlet superconducting pairing on the nearest-neighbor bonds along the directions ${\bm R}_{j}$ shown in~\fig{fig:graphenelattice}. This gives the pairing term
\beq 
  \Delta_{\pm}({\bm k}) = 
  \Delta\sum_{j=1}^3 e^{\mp i\phi_j}  \left(\begin{array}{cc} 0 &
  e^{i{\bm k}\cdot{\bm R}_j} \\
  e^{-i{\bm k}\cdot{\bm R}_j} & 0 \end{array}\right).  \label{eq:gapmatrix}
\eeq
The magnitude $\Delta$ of the pairing potential 
is the same for each bond $j$, but the phase is  $\phi_j=(j-1)2\pi/3$, 
and the two choices of sign in~\eq{eq:gapmatrix} define degenerate 
pairing potentials with opposite chiralities.  A similar chiral spin-triplet pairing is discussed in Appendix \ref{sec:chiralp}.

The two pairing
potentials in \eq{eq:gapmatrix} can be written in terms of the Pauli
matrices $s_\lambda$  
\beqarray
  && \Delta_+(\bm k) = \Delta^x_{\bm k}s_x + \Delta^y_{\bm k}s_y \,, 
  \label{eq:D+} \\
  && \Delta_-(\bm k) = (\Delta^x_{\bm k})^*s_x + (\Delta^y_{\bm k})^*s_y \,, 
  \label{eq:D-}
\eeqarray
where
\beqarray
  && \Delta^x_{\bm k} = \Delta\sum_{j=1}^3 e^{-i\phi_j} \cos({\bm k}\cdot{\bm R}_j) \,, 
\label{eq:Dx} \\
  && \Delta^y_{\bm k} = -\Delta\sum_{j=1}^3 e^{-i\phi_j} \sin({\bm k}\cdot{\bm R}_j) \,.
\label{eq:Dy} 
\eeqarray  %
The pairing potentials in \eq{eq:gapmatrix} can also be expressed in
terms of basis states of the irreducible representation
$E_{2g}$  
\beq
\Delta_{\pm}({\bm k}) = \Delta_{x^2-y^2}({\bm k}) \pm
i\Delta_{xy}({\bm k})\,, 
\eeq
where
\beqarray
\Delta_{x^2-y^2}({\bm k})& = &
\Delta\left\{\left[\cos(k_xa) -
  \cos(\tfrac{1}{2}k_xa)\cos(\tfrac{\sqrt{3}}{2}k_ya)\right]{s}_x\right. \notag
\\
&& \left.-  \left[\sin(k_xa) +
  \sin(\tfrac{1}{2}k_xa)\cos(\tfrac{\sqrt{3}}{2}k_ya)\right]
                  {s}_y\right\}, \notag \\ 
\Delta_{xy}({\bm k}) & = &
    \sqrt{3}\Delta\left\{-\sin(\tfrac{1}{2}k_xa)\sin(\tfrac{\sqrt{3}}{2}k_ya){s}_x
    \right. \notag \\
    && \left .+  \cos(\tfrac{1}{2}k_xa)\sin(\tfrac{\sqrt{3}}{2}k_ya)
{s}_y\right\}. \notag 
\eeqarray
 and $a$ is the distance between neighboring sites.
When projected onto the states near the Fermi surface, the basis
states $\Delta_{x^2-y^2}({\bm k})$ and $\Delta_{xy}({\bm k})$ 
have the forms of $d_{x^2-y^2}$ and $d_{xy}$ waves,
so $\Delta_{\pm}({\bm k})$ can be regarded as a chiral
$(d_{x^2-y^2}\pm id_{xy})$-wave pairing state. 
The matrices ${s}_x$ and ${s}_y$ are multiplied by the functions
that are even and odd with respect to ${\bm k}\to-{\bm k}$.
This ensures that the pairing potentials are even under inversion, 
e.g.\ ${\cal I}^\dagger\Delta_{x^2-y^2}({\bm k}){\cal I} 
= \Delta_{x^2-y^2}(-{\bm k})$, as the inversion operator ${\cal I}=s_x$ 
swaps the sublattice index.  A similar sublattice gap structure has 
been proposed for the chiral $f$-wave pairing state
in UPt$_3$.~\cite{Yanase2016,KerrUPt3_theory}

%%%%%%%%%%%%%%%%%%%%%%%%%%%%%%%%%%%%%%%%%%%%%%%%
\section{Time-reversal-odd bilinear}
\label{sec:TROB-hex}
%%%%%%%%%%%%%%%%%%%%%%%%%%%%%%%%%%%%%%%%%%%%%%%%

A central goal of our work is to understand how broken
time-reversal symmetry in the particle-particle superconducting
channel can lead to observable effects in the particle-hole
channel, e.g.,\ the anomalous Hall conductivity and the
polar Kerr effect.  For such effects, it is not sufficient to consider the 
pairing potential $\Delta(\bm k)$ alone, since it is not gauge-invariant.  
Rather, these observable must depend on a time-reversal
symmetry-breaking bilinear combination of $\Delta(\bm k)$ and $\Delta^\dagger(\bm k)$.

In order to define the time-reversal operation, let us label the second-quantized electron operators $\psi_{c,\sigma}(\bm k)$ by the sublattice index $c$ and the spin index $\sigma=\pm$.  The time-reversal operation involves the substitution 
$\psi_{c,\sigma}(\bm k)\to\sigma\psi_{c,-\sigma}(-\bm k)$ and complex conjugation of the matrix elements in the BdG Hamiltonian.~\cite{Bernevig}
Then the off-diagonal term in \eq{eq:HBdG} transforms as follows
  \beqarray  
  \psi^\dagger_{c,\downarrow}(\bm k) \Delta_{cd}(\bm k)
  \psi^\dagger_{d,\uparrow}(-\bm k) &\to &
  -\psi^\dagger_{c,\uparrow}(-\bm k) \Delta_{cd}^*(\bm k)
  \psi^\dagger_{d,\downarrow}(\bm k) \notag \\
  &
  =& \psi^\dagger_{c,\downarrow}(\bm k) \Delta_{dc}^*(\bm k) 
  \psi^\dagger_{d,\uparrow}(-\bm k)  \label{TR-Delta}
  \eeqarray
where summation over repeated indices is implied.   Note that, to
  obtain the second line, we anticommuted the fermion operators and then swapped
  the  sublattice indices.  Thus we obtain a BdG 
Hamiltonian of the same form with  
\beq  \label{D->D+}
\Delta(\bm k) \;\to\; \Delta^\dagger(\bm k)
\eeq
upon time reversal.

The simplest bilinear product of the pairing potential with its
Hermitian conjugate is $\Delta(\bm k)\Delta^\dagger(\bm k)$.\cite{bilinear}  
The time-reversal-odd part of this bilinear product, which we abbreviate 
as $\mbox{TROB}$, 
obtained as the difference between $\Delta(\bm k)\Delta^\dagger(\bm k)$ 
and its time-reversed counterpart, is a commutator
  \beq \label{eq:TROB}
  \mbox{TROB} = \Delta(\bm k)\Delta^\dagger(\bm k)
  -\Delta^\dagger(\bm k)\Delta(\bm k)  = [\Delta(\bm k),\Delta^\dagger(\bm k)].  
  \eeq
Due to its gauge invariance and odd time-reversal behavior, a non-zero
$\mbox{TROB}$ permits broken time-reversal symmetry in the
particle-particle channel to manifest in the particle-hole channel.
In Appendix~\ref{sec:TROBgeneral}, we show that the expression for the
$\mbox{TROB}$ in \eq{eq:TROB} applies to more general Hamiltonians,
which may include spin-orbit coupling and more electronic degrees of
freedom, or break inversion symmetry.
In the second-quantized formalism, the TROB matrix from \eq{eq:TROB} appears in the time-reversal-odd part of the commutator of the pairing terms
  \beqarray  \notag
  && [\hat H_{\Delta},\hat H_{\Delta}^\dagger] 
  - \Theta [\hat H_{\Delta},\hat H_{\Delta}^\dagger] \Theta^{-1} \\
  && =\sum_{\bm k,\sigma} \psi^\dagger_{c,\sigma}(\bm k) \, 
  [\Delta(\bm k),\Delta^\dagger(\bm k)]_{cd} \, \psi_{d,\sigma}(\bm k),
  \label{eq:commutator}
  \eeqarray
where $\Theta$ is the time-reversal operation, and
  \beq  \label{eq:H_Delta}
  \hat H_{\Delta}=\sum_{\bm k} \psi^\dagger_{c,\downarrow}(\bm k) \Delta_{cd}(\bm k)
  \psi^\dagger_{d,\uparrow}(-\bm k).
  \eeq

We immediately see that the $\mbox{TROB}$ in \eq{eq:TROB}
always vanishes for a single-band spin-singlet superconductor
where $\Delta(\bm k)$ is just a complex function.  This implies that any
probe of time-reversal symmetry breaking, e.g.,\ the Hall conductivity or polar Kerr effect, must vanish if such a system is clean, 
so that the momentum $\bm k$ is a good quantum number.
Hence, the experimental detection of time-reversal symmetry breaking 
in single-band superconductors must rely upon 
inhomogeneities  not conserving $\bm k$, e.g.,\ 
scattering off impurities.~\cite{Lutchyn2009}

However, for a clean multiband system, where the pairing potential can be expressed as a matrix in the band indices, it is 
possible for the commutator in~\eq{eq:TROB} to take on nonzero
values.  This is the case for the honeycomb lattice model
of Sec.~\ref{sec:model}, for which we obtain
\begin{align}  % \lefteqn{}
  & \mbox{TROB}
  =[\Delta_\pm({\bm k}),\Delta^\dagger_\pm({\bm k})]
  =\pm 2 i[\bm\Delta_{\bm k}\wedge\bm\Delta_{\bm k}^*]s_z 
  \notag \\
 &  =\pm 4 s_z |\Delta|^2 \sum_{j<j'} 
 \sin(\phi_j-\phi_{j'}) \, \sin[\bm k\cdot(\bm R_j-\bm R_{j'})] 
\notag \\
 & = \pm 4 \sqrt{3} |\Delta|^2
\sin(\tfrac{\sqrt{3}}{2}k_ya)\left[\cos(\tfrac{3}{2}k_xa)-\cos(\tfrac{\sqrt{3}}{2}k_ya)\right]
s_z, \notag\\
 \label{eq:gapprod}
\end{align}
where the wedge product $[\bm a\wedge\bm b]=a_xb_y-a_yb_x$ 
is used for the two-component vector 
$\bm\Delta_{\bm k}=(\Delta^x_{\bm k},\Delta^y_{\bm k})$
from Eqs.~(\ref{eq:Dx}) and (\ref{eq:Dy}). 
In the second line, the sum is taken over the pairs of nearest-neighbor bonds in Fig.~\ref{fig:graphenelattice}.  The nonzero TROB  in \eq{eq:gapprod} implies the existence of a time-reversal symmetry-breaking sublattice polarization of the pairing state, which we define as
\begin{align}
  \Xi_\pm({\bm k}) &=\mbox{Tr}\{\Delta^\dagger_{\pm}({\bm
    k}){s}_z\Delta_{\pm}({\bm k})\} \notag \\
  & =\pm 4\sqrt{3}|\Delta|^2
  \sin(\tfrac{\sqrt{3}}{2}k_ya)\left[\cos(\tfrac{3}{2}k_xa)-\cos(\tfrac{\sqrt{3}}{2}k_ya)\right]. \label{eq:polarization}
\end{align}
The sublattice polarization $\Xi_\pm(\bm k)$ is crucially important
for the physical effects discussed in the rest of the paper. It
quantifies the relative participation in the pairing of electrons on
the $A$ and $B$ sites. Pairing at the $M$ point of the Brillouin zone
involves both sublattices equally, and so $\Xi_\pm({\bm k}_M)=0$. In
contrast, pairing  
at the $K$ ($K'$) point involves exclusively the $B$ ($A$)
sublattice for the $\Delta_+$ potential, and so $\Xi_+({\bm
  k}_K)=-\Xi_{+}({\bm k}_{K'})=9|\Delta|^2$; the sublattice polarization is 
reversed for $\Delta_-$.~\cite{QiFuSunGu2017}  This can be considered as a
generalization to non-spin internal degrees of freedom of the spin
polarization of a single-band nonunitary triplet state.~\cite{SigUeda1991} 
It has recently been pointed out that such a
polarization generically arises in multiband time-reversal
symmetry-breaking superconductors, where it can have dramatic effects
on the low-energy nodal structure.~\cite{4x4} Although the
effect of the polarization on the electronic structure is 
confined to high energies in our fully-gapped pairing state, we shall see
below that it plays a key role in generating the Hall conductivity.

Further insight into the implications of a nonzero TROB is provided by the
concept of the superconducting fitness, which has recently emerged as
a way to characterize the pairing state in multiband
materials.\cite{Ramires2016,Ramires2018} For our system, where the
normal-state Hamiltonian $H_0(\bm k)$ is time-reversal invariant, a
superconducting state is said to have perfect fitness when
\beq \label{eq:BestFitCommut} 
[H_0(\bm k),\Delta(\bm k)]=0,
\eeq
i.e.,\ the normal-state Hamiltonian $H_0(\bm k)$ commutes with the pairing
potential $\Delta(\bm k)$.  Then these two matrices can be simultaneously diagonalized in the normal-state band basis, and so there is no
interband pairing in the case of perfect fitness. 
In this basis, a multiband BdG Hamiltonian with
even-parity spin-singlet pairing reduces to 
a collection of decoupled single-band terms, 
so the $\mbox{TROB}$ must therefore vanish.
We conclude that the lack of perfect fitness, i.e.,\ a violation of
\eq{eq:BestFitCommut} and the presence of interband pairing, 
is a necessary (but not sufficient) condition for a nonvanishing TROB. The presence of interband pairing has been previously
noted as crucial for the existence of the polar Kerr effect in
clean chiral superconductors.~\cite{Wysokinski2012,Taylor2012}

The chiral $d$-wave pairing potential in our model does violate the superconducting fitness condition:
\beq  \label{eq:+-Fit}
  [H_{0}({\bm k}),\Delta_\pm({\bm k})]
  =  2i[\bm\epsilon_{\bm k}\wedge\bm\Delta_{\bm k}^{(*)}]s_z \neq 0,
\eeq
where $\bm\epsilon_{\bm k}=(\epsilon^x_{\bm k},\epsilon^y_{\bm k})$, complex conjugation in the right-hand side applies only for the negative chirality, and we set the Semenoff term to zero for simplicity (i.e.,\ $\delta_s=0$).  The violation of the fitness condition is due to the nontrivial phases $\phi_j$ along the
nearest-neighbor bonds, making the complex vector $\Delta_{\bm k}$ in 
Eqs.~(\ref{eq:Dx}) and (\ref{eq:Dy}) not parallel to the real vector $\bm\epsilon_{\bm k}$ in \eq{eq:epsilon-xy}.  Although the presence of both intraband and interband pairing in the chiral $d$-wave state is energetically disadvantageous due to mismatch of the energies of different
bands,\cite{Mineev2012} it can emerge in a mean-field BCS theory due
to the short range of real-space interaction between electrons. 
Indeed, the pairing potential~\eq{eq:gapmatrix} naturally arises from 
a mean-field decoupling of the nearest-neighbor exchange interactions 
in a $t$-$J$ model.~\cite{BlaSchHonJPCM2014}

It is instructive to compare our results to a chiral $d$-wave state
with purely intra-sublattice (i.e.,\ next-nearest-neighbor) pairing,
as proposed in~\Ref{Kiesel2012}.  For this state, 
the pairing potential
\begin{align} \label{eq:chirald_intra}
\tilde{\Delta}_\pm(\bm k) =&
\tilde{\Delta}\left[\cos(\sqrt{3}k_ya)-\cos(\tfrac{3}{2}k_xa)\cos(\tfrac{\sqrt{3}}{2}k_ya)\right.\notag \\
  & \left.\pm i\sqrt{3}\sin(\tfrac{3}{2}k_xa)\sin(\tfrac{\sqrt{3}}{2}k_ya)\right]s_0
\end{align}
is proportional to the unit matrix in sublattice space. As
such, this potential commutes with the normal-state Hamiltonian and so
has perfect fitness. Thus, despite the fact that
$\tilde{\Delta}_\pm(\bm k)$ breaks time-reversal symmetry and 
has a nonzero phase winding around the Fermi surface, 
this state does not display an intrinsic polar Kerr effect 
because $\mbox{TROB}=0$.

The pairing potential $\Delta_\pm({\bm k})$ and the
TROB describe spontaneous breaking of the continuous
$U(1)$ gauge symmetry and the discrete  
$\mathbb{Z}_2$ time-reversal symmetry, respectively.  In the
mean-field BCS theory, both symmetries are broken simultaneously.
In a more general framework, however, these 
two symmetries may be broken at separate phase transitions taking
place at different temperatures.  For example, the TROB 
may acquire a non-zero expectation  value at a higher
temperature by selecting positive or negative chirality (which can be
detected experimentally by observing the polar Kerr effect), while the
expectation value of the pairing potential $\Delta_\pm({\bm k})$ is
still zero due to phase fluctuations.  The superconducting properties,
such as supercurrent and Meissner effect, would emerge at a lower
temperature, where $\Delta_\pm({\bm k})$ acquires a non-zero
expectation value.  This scenario is discussed in more detail in
Sec.~\ref{sec:phenomeno}.

The above considerations are not limited to  spin-singlet even-parity
superconductivity.  In Appendix~\ref{sec:chiralp}, we show that an odd-parity spin-triplet chiral $p$-wave pairing has a similar TROB and sublattice polarization.

%%%%%%%%%%%%%%%%%%%%%%%%%%%%%%%%%%%%%%%%%%%%%%%%
\begin{figure*}
  (a) \includegraphics[height=0.75\columnwidth]{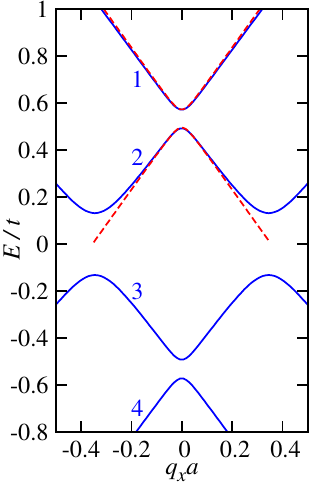} \hspace*{1ex}
  (b)
  \includegraphics[height=0.75\columnwidth]{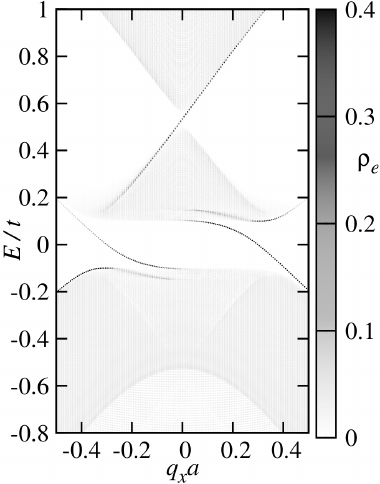} \hspace*{1ex}(c)
  \includegraphics[height=0.75\columnwidth]{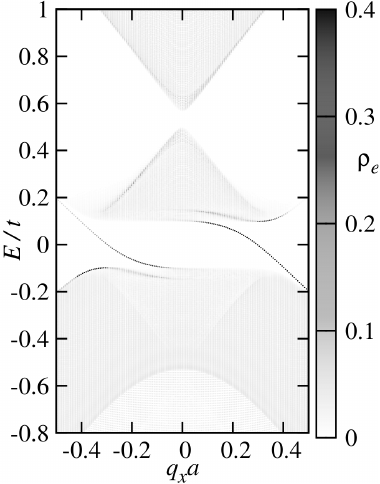}
\caption{(Color online) (a) 
  The eigenvalues of the BdG Hamiltonian in~\eq{eq:HBdG} (solid blue lines)
  and the effective gapped Dirac model in \eq{eq:H-eff} 
  (dashed red lines)  vs.\ the deviation from 
  the $K$ point, ${\bm k}_K = (0,4\pi/3\sqrt{3}a)$.
  (b) Energy spectrum obtained by exact diagonalization
  for a slab of 1200 layers in the $y$ direction and armchair edge. The
  eigenvalues are 
  shaded according to the integrated electron-like weight in the
  first 40 laters of the slab. A dispersing edge resonance is clearly
  visible within the gap at the Dirac point near $E\approx0.5t$.
  Two dispersing Andreev edge states are also present inside the
  superconducting gap near $E=0$. (c) Same as in panel (b),
  but with the inclusion of a Semenoff term $\delta_s=0.065t$, which
  renders the gap at the Dirac point topologically trivial. In
  all plots we set $\Delta=0.0866t$ and $\mu=-0.5t$.} 
  \label{fig:gappeddispersion}
\end{figure*}
%%%%%%%%%%%%%%%%%%%%%%%%%%%%%%%%%%%%%%%%%%%%%%%%

%%%%%%%%%%%%%%%%%%%%%%%%%%%%%%%%%%%%%%%%%%%%%%%%
\section{Topological mass gap}\label{sec:massgap}
%%%%%%%%%%%%%%%%%%%%%%%%%%%%%%%%%%%%%%%%%%%%%%%%

Let us set the Semenoff term in \eq{eq:H0k} to zero first:
$\delta_s=0$.  In this case, there is no gap at the Dirac points $K$ 
and $K'$ in the normal state. However, the energy spectrum of the BdG
Hamiltonian in \eq{eq:HBdG} shows an unexpected gap opening at the Dirac
points near $E=\pm\mu=\pm0.5t$ in~\fig{fig:gappeddispersion}(a), far
away from  the usual superconducting gap at the Fermi level $E=0$.
Note that the
momentum ${\bm q}={\bm k}-{\bm k}_{K}$ is measured relative to the $K$ 
point in \fig{fig:gappeddispersion}.

To gain insight into the nature of this unexpected gap, we derive an
effective Hamiltonian for the states near the Dirac points, 
perturbatively including the superconducting pairing in the limit
$\Delta\ll|\mu|$. Our starting point is the formal expression for the
electron-like component of the Green's function,
\begin{align} \label{eq:G-1}
  \lefteqn{\hbar G^{-1}({\bm k},\omega) }\notag \\
    & = \hbar \omega - H_0({\bm k}) -
\Delta({\bm k})\left[\hbar \omega + H_0^T(-{\bm k})\right]^{-1}\Delta^\dagger({\bm k})\,.
\end{align}
To find the energy spectrum in the vicinity of the Dirac points, we replace
$\omega \to -\mu$ in the last term of
\eq{eq:G-1} and obtain an effective Hamiltonian 
\beq  \label{eq:H-eff}
H_{\text{eff}}({\bm k}) \approx H_0({\bm k}) + \delta H\,.
\eeq
Near the $K$ point, we can expand  the first term 
to linear order in the relative momentum ${\bm q}={\bm k}-{\bm k}_K$
\beq  \label{eq:H0q}
H_{0}({\bm q}) = \left(\begin{array}{cc}
-\mu & \frac{3}{2}ta(- q_y-iq_x)\\
\frac{3}{2}ta(- q_y+iq_x)& -\mu
\end{array}\right)\,,
\eeq
with the correction due to superconductivity 
\beq
\label{eq:delta-HK}
\delta H ({\bm k}_K)  
= -\frac{\Delta_{\pm}({\bm k}_K)\Delta^\dagger_{\pm}({\bm k}_K)}{2\mu} 
= -\frac{9|\Delta|^2}{4\mu}({s}_0\mp{s}_z)\,.
\eeq
Near the $K'$ point, the expansion of the unperturbed Hamiltonian is
identical except for the reversed sign in front of $q_y$, 
and the correction is
\beq  \label{eq:delta-HKp}
\delta H({\bm k}_{K'})
= -\frac{\Delta_{\pm}({\bm k}_{K'})\Delta^\dagger_{\pm}({\bm k}_{K'})}{2\mu} 
= -\frac{9|\Delta|^2}{4\mu}({s}_0\pm{s}_z)\,.
\eeq
Note that Eqs.\ (\ref{eq:delta-HK}) and (\ref{eq:delta-HKp})
can be obtained from the last term in \eq{eq:G-1} only in the vicinity
of the Dirac points, where $H_{0}$ in \eq{eq:H0q} is proportional to the unit
matrix in the limit of vanishing ${\bm q}$.  Equation (\ref{eq:H-eff})
can be interpreted as an effective normal-state Hamiltonian with the
second-order perturbative correction due to superconducting pairing. 

The perturbative correction given by Eqs.~(\ref{eq:delta-HK})
and (\ref{eq:delta-HKp}) is proportional to the matrix product
$\Delta({\bm k})\Delta^\dagger({\bm k})$.  
Its time-reversal-even part,
proportional to the unit matrix $s_0$, shifts the energy of the Dirac
point. In contrast, the time-reversal-odd part (i.e.,\ the TROB),
proportional to $s_z$, opens a mass gap. This demonstrates 
the appearance of time-reversal symmetry breaking in the particle-hole
channel due to the nonzero TROB.
The gapped energy dispersion derived via this 
perturbative argument, 
shown by the dashed red curve in~\fig{fig:gappeddispersion}(a), 
is in excellent agreement with the dispersion of the full model near
to the Dirac point.

%%%%%%%%%%%%%%%%%%%%%%%%%%%%%%%%%%%%%%%%%%%%%%%%
\begin{figure}[b]
\includegraphics[width=0.9\columnwidth,clip]{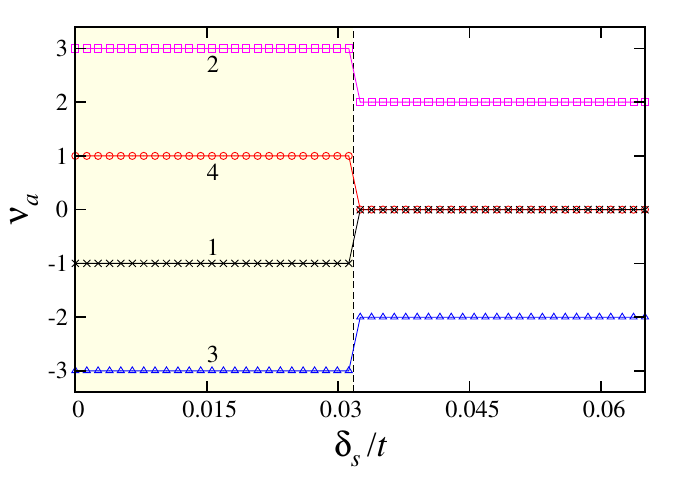}
\caption{(Color online) Variation of the Chern numbers
  $\nu_\alpha$ of the four BdG bands labelled as
  in~\fig{fig:gappeddispersion}(a)  vs.\ the Semenoff  
  term $\delta_s$.   The shaded region $\delta_s<\delta_c$ 
  corresponds to the topologically nontrivial state.
  The lines are a guide for the eye. Other
  parameters are the same as
  in~\fig{fig:gappeddispersion}.}~\label{fig:Chern} 
\end{figure}
%%%%%%%%%%%%%%%%%%%%%%%%%%%%%%%%%%%%%%%%%%%%%%%%

The mass gaps at the $K$ and $K'$ points introduced by the
superconductivity [Eqs.~(\ref{eq:delta-HK}) and~(\ref{eq:delta-HKp})]
have opposite signs. This suggests a topologically nontrivial state,  
as in Haldane's model of the quantum anomalous Hall state on the
honeycomb lattice.~\cite{Haldane1988} The topological nature of the
mass gap is confirmed by calculation of the Chern numbers for the
different bands and observation of chiral edge states within the
energy gaps via the bulk-boundary correspondence. With the opening of
the mass gap, the four eigenstates of the BdG 
Hamiltonian in \eq{eq:HBdG} are everywhere nondegenerate, so 
a Chern number $\nu_\alpha$ can be defined for each band
$\alpha=1,2,3,4$, as labeled in~\fig{fig:gappeddispersion}(a).
As shown in~\fig{fig:Chern}, each band has a nonzero Chern number
for $\delta_s=0$, i.e.\ in the absence of the Semenoff term.
The sum of the Chern numbers of the occupied bands 3 and 4 below the
chemical potential is $-2$, consistent with the chiral $d$-wave
superconductivity. Correspondingly, the two topologically-protected
chiral edge states within the superconducting gap are clearly visible
near $E=0$ in the energy spectrum weighted by the integrated probability density
of the electron-like wave
function components near the surface, as shown
in~\fig{fig:gappeddispersion}(b) for the 
armchair edge. The nonzero Chern numbers of the outer bands 1 and 4,
which are separated by the mass gap from the inner  bands 2 and 3,
imply that the mass gap is topological. Thus, we would expect to find
a single chiral edge state within each mass gap.  However, due to the
spectrum doubling in the superconducting state,  the hole-like states 
generally overlap with the energy range of the mass gap and can
hybridize with the edge state. Nevertheless, the edge state persists
as a predominantly electron-like {\it  edge resonance} inside the mass
gap between bands 1 and 2 in~\fig{fig:gappeddispersion}(b).  

A combination of a nonzero Semenoff term $\delta_s \neq0$ 
in \eq{eq:H0k} and the superconducting corrections in Eqs.\ (\ref{eq:delta-HK})
and (\ref{eq:delta-HKp}) produces different magnitudes of the mass gaps 
at the two Dirac points $K$ and $K'$. At a critical value
$\delta_c=\frac{1}{2}(\sqrt{9|\Delta|^2+\mu^2}-|\mu|)$, the gap
at one of the Dirac points passes through zero and changes sign.
Correspondingly, as shown in~\fig{fig:Chern}, there is an
abrupt change in the Chern numbers of all BdG bands at this
topological phase transition. For $|\delta_s|>\delta_c$, the Chern
number of the outer bands 1 and 4 vanishes, although the sum of the
Chern numbers of the occupied bands 3 and 4 remains $-2$. This is
consistent with the mass gaps at $K$ and $K'$ having the same sign,
which is topologically trivial. Accordingly, we do not observe any
edge resonance within the gap, as shown
in~\fig{fig:gappeddispersion}(c). 

Repeating the calculations for a zigzag edge, we also find evidence
for Haldane states. However, they are mixed with the standard
flat-band edge states that exist at the zigzag edges of a hexagonal 
lattice, making their interpretation more complicated.

%%%%%%%%%%%%%%%%%%%%%%%%%%%%%%%%%%%%%%%%%%%%%%%%
\section{Loop currents}\label{sec:loops}
%%%%%%%%%%%%%%%%%%%%%%%%%%%%%%%%%%%%%%%%%%%%%%%%

\begin{figure}
\includegraphics[height=0.4\columnwidth]{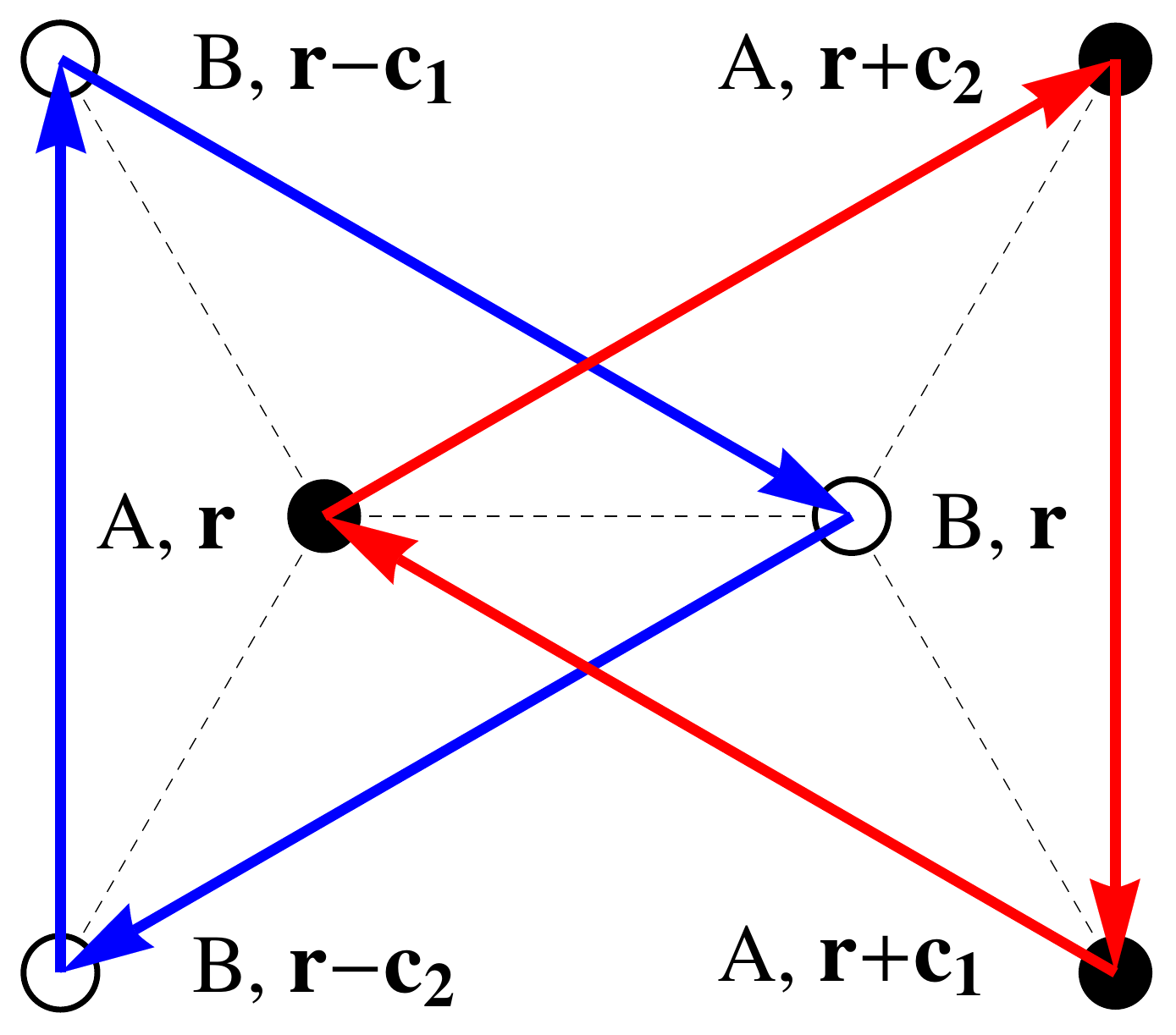}
\caption{(Color online) 
  Loop currents between the three 
  nearest neighbors of the $A$ and $B$ sites in 
  the unit cell at $\bm r$.}
  \label{fig:effectivehopping} 
\end{figure}

It was argued in the previous  Section that the energy gaps observed
at the Dirac points are similar to the energy gaps in Haldane's model
of the quantum anomalous Hall insulator.~\cite{Haldane1988} They arise
in Haldane's model due to the presence of a time-reversal symmetry-breaking
next-nearest-neighbor hopping  term, resulting in loop currents 
around each lattice site shown by the arrows in~\fig{fig:effectivehopping}.  
In second quantization, the time-reversal-odd part of this hopping
term is proportional to the dimensionless operator $\chi_{lc}$, which
is defined as  
\beqarray 
  \chi_{lc} & = &  i\sum_{{\bm r},\sigma} \left( a^\dagger_{{\bm r},\sigma}a_{{\bm
      r}+{\bm c}_1,\sigma} + a^\dagger_{{\bm r}+{\bm c}_1,\sigma}a_{{\bm
      r}+{\bm c}_2,\sigma} + a^\dagger_{{\bm r}+{\bm c}_2,\sigma}a_{{\bm
      r},\sigma} \right. \notag \\
  & &\left. + b^\dagger_{{\bm r},\sigma}b_{{\bm
      r}-{\bm c}_1,\sigma} + b^\dagger_{{\bm r}-{\bm c}_1,\sigma}b_{{\bm
      r}-{\bm c}_2,\sigma} + b^\dagger_{{\bm r}-{\bm c}_2,\sigma}b_{{\bm
      r},\sigma} -\Hc \right) \notag \\
  & = & \sum_{{\bm
    k}}4\sin(\tfrac{\sqrt{3}}{2}k_ya)\left[\cos(\tfrac{3}{2}k_xa)-
    \cos(\tfrac{\sqrt{3}}{2}k_ya)\right] \notag \\
  && \times \Psi^\dagger_{\bm
    k}{\tau}_0\otimes{s}_z\Psi_{\bm k}\,. \label{eq:loopcurrent}
  \eeqarray
  Here ${\bm c}_1$ and ${\bm c}_2$ are the primitive lattice vectors
(see~\fig{fig:graphenelattice}), the operator $a_{{\bm r},\sigma}$
($b_{{\bm r},\sigma}$) destroys a spin-$\sigma$ electron on the $A$
($B$) site of the unit cell corresponding to the lattice vector ${\bm
  r}$, and ${\tau}_0$ is the unit matrix in Nambu space.  The
  sign convention in \eq{eq:loopcurrent} matches the convention for
  the link directions in \fig{fig:effectivehopping}.

%%%%%%%%%%%%%%%%%%%%%%%%%%%%%%%%%%%%%%%%%%%%%%%%
%%%%%%%%%%%%%%%%%%%%%%%%%%%%%%%%%%%%%%%%%%%%%%%%

In Haldane's model, the operator in \eq{eq:loopcurrent}
has a nonzero expectation value $\langle\chi_{lc}\rangle\neq0$,
resulting in the loop currents shown in~\fig{fig:effectivehopping}.
In our model, the operator $\chi_{lc}$ appears in the commutator (\ref{eq:commutator}) 
with the TROB given by \eq{eq:gapprod}.  The commutator of the pairing
terms on the adjacent nearest-neighboring links generates electron
transfer between the next-nearest neighboring sites with a complex
amplitude carrying the phase difference of pairing potentials shown in
Fig.~\ref{fig:graphenelattice}. The 
analogy between our system and Haldane's model
suggests that $\chi_{lc}$ 
also has a non-zero expectation value in the
chiral $d$-wave state, which is readily verified to be 
\beqarray
\frac{\langle \chi_{lc} \rangle}{N} &=& -\frac{1}{N}\sum_{{\bm
    k}}\sin(\tfrac{\sqrt{3}}{2}k_ya)\left[\cos(\tfrac{3}{2}k_xa) -
  \cos(\tfrac{\sqrt{3}}{2}k_ya)\right] \notag \\
&& \times \frac{1}{\beta}\sum_{\nu_m}
\frac{8\mu\, \Xi_\pm({\bm k}) 
  }{(\hbar^2\nu_m^2 + E_{{\bm k},1}^2)(\hbar^2\nu_m^2 + E_{{\bm k},2}^2)}\,. 
  \label{eq:lc_expectation} 
\eeqarray
Here $E_{{\bm k}, \alpha=1,2}>0$ are the quasiparticle dispersions
corresponding to the upper two bands shown
in~\fig{fig:gappeddispersion}(a) (explicit expressions are given
in~\Ref{BlaSchPRB2014}), $\beta=1/k_BT$ is the inverse temperature, 
$\nu_m$ are the fermionic Matsubara
frequencies, and $N$ is the number of unit cells.  The  
presence  in \eq{eq:lc_expectation} of the sublattice
  polarization $\Xi_\pm({\bm k})$ from \eq{eq:polarization}
  (equivalent to the TROB) 
is essential for obtaining a nonzero expectation value of
  $\chi_{\text{lc}}$. 
 Since $\Xi_\pm({\bm k})$ has the same
momentum dependence  
as the term in front of the fraction in \eq{eq:lc_expectation}, 
the summand has the same sign everywhere in the Brillouin zone, 
and thus the expectation value $\langle\chi_{lc}\rangle$ is 
  nonzero. The essential importance of the TROB in ensuring 
$\langle\chi_{lc}\rangle\neq0$ is consistent with the role of
the TROB in generating the  energy 
gaps at the Dirac points.
As such, the inclusion of the Semenoff term does not alter the conditions 
for a nonzero expectation value of $\langle\chi_{lc}\rangle$, but
\eq{eq:lc_expectation} is replaced by a more complicated expression  
given in Appendix~\ref{sec:genexp}. We note that 
$\langle\chi_{lc}\rangle\neq0$ was calculated in~\Ref{Faye2015}
for the closely-related nearest-neighbor chiral $p$-wave state
introduced in Appendix~\ref{sec:chiralp}. In contrast,
$\langle\chi_{lc}\rangle$ is zero for the intra-sublattice chiral $d$-wave state described by
\eq{eq:chirald_intra}, where the TROB vanishes.

However, unlike in Haldane's model, a nonzero expectation value of
$\chi_{lc}$ in our system  
only implies the presence of loop current {\it correlations}.
Since the normal-state Hamiltonian (\ref{eq:H0k}) does not contain
next-nearest-neighbor hopping, 
 there are no  
current operators 
between next-nearest sites in our model.
 We can remedy this by introducing a next-nearest-neighbor hopping term 
with  a small 
 real amplitude $t'$: 
\beq \label{t'}
H_{t'}=t'\sum_{\langle\langle m,n
  \rangle\rangle,\sigma}(a_{{\bm r}_m,\sigma}^\dagger a_{{\bm r}_n,\sigma} +
b_{{\bm r}_m,\sigma}^\dagger b_{{\bm r}_n,\sigma}). 
\eeq
The  corresponding current operators between the next-nearest-neighbor sites 
$m$ and $n$  belonging to the sublattice $ c=A,B$  are   
hence  obtained from \eq{t'} as
\beq \label{Inm}
I^c_{mn}=\frac{iet'}{\hbar}\left[\psi^\dagger_{c,\sigma}({\bm
    r}_m)\psi^{}_{c,\sigma}({\bm r}_n) - \psi^\dagger_{c,\sigma}({\bm
    r}_n)\psi^{}_{c,\sigma}({\bm r}_m)\right],
\eeq
where $\psi^{}_{c,\sigma}({\bm r})$ is the annihilation
operator for spin-$\sigma$ electrons on sublattice $c$ of unit cell ${\bf r}$.
 Adding the 
 current operators  in \eq{Inm} with  the signs 
corresponding to \fig{fig:effectivehopping}, we introduce 
the total current operator $I_{\rm tot}$ as 
\beq \label{Itot}
I_{\rm tot}=\frac{et'}{\hbar}\chi_{lc}\,.
\eeq
 Then the expectation value $I$ 
 of the microscopic current on one link is obtained as
\beq \label{I} 
I=\frac{\langle I_{\rm tot}\rangle}{6N}=\frac{et'}{6\hbar}\frac{\langle\chi_{lc}\rangle}{N},
\eeq
 where we divide by $6$ because there are six currents of equal
 magnitude in a unit cell. 
 The current $I$ is very small, because it is proportional to the small hopping amplitude $t'$ and $\langle\chi_{lc}\rangle/N\sim\mu\Delta^2/t^3\ll 1$ from \eq{eq:lc_expectation}.

 Another physical consequence of $\langle\chi_{lc}\rangle\neq0$ is the
existence of a nonzero anomalous Hall conductivity in the absence of
an external magnetic field, which is calculated in the next Section.
Unlike the current $I$ in \eq{I}, the Hall conductivity
does not require $t'\neq0$, so we set $t'=0$ in the rest of the paper to
simplify calculations.\\

%%%%%%%%%%%%%%%%%%%%%%%%%%%%%%%%%%%%%%%%%%%%%%%%
\section{Hall conductivity}\label{sec:hall}
%%%%%%%%%%%%%%%%%%%%%%%%%%%%%%%%%%%%%%%%%%%%%%%%

The existence of loop-current correlations in
\eq{eq:lc_expectation}  for 
the chiral $d$-wave state naturally suggests the 
presence of an intrinsic Hall conductivity. Indeed, the nontrivial
sublattice structure of the BdG Hamiltonian (\ref{eq:HBdG}) is
consistent with the conditions outlined in~\Ref{Taylor2012} for the  
existence of an intrinsic Hall effect.

\begin{figure}
\includegraphics[width=0.85\columnwidth,clip]{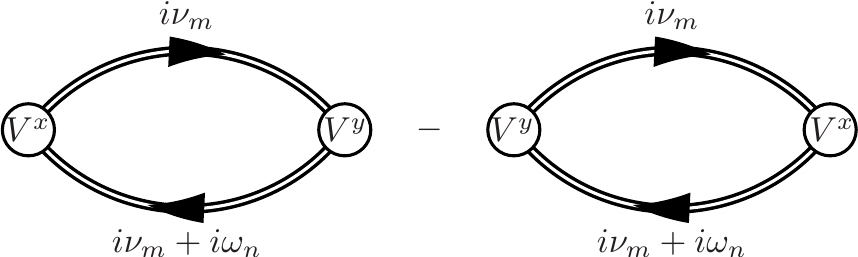}
\caption{Feynman diagrams for the intrinsic Hall conductivity. 
  The double lines are the  matrix 
  superconducting Green's functions  in Nambu notation. 
  The vertices ${V}^x$ and ${ V}^y$ are the $x$- and
  $y$-components of the velocity operator in~\eq{eq:V}.
  The Matsubara frequencies $i\omega_n$ and $i\nu_m$ are the external 
  bosonic and internal fermionic frequencies, respectively.}~\label{fig:bubble}
\end{figure}

\begin{figure}[b]
\includegraphics[width=\columnwidth]{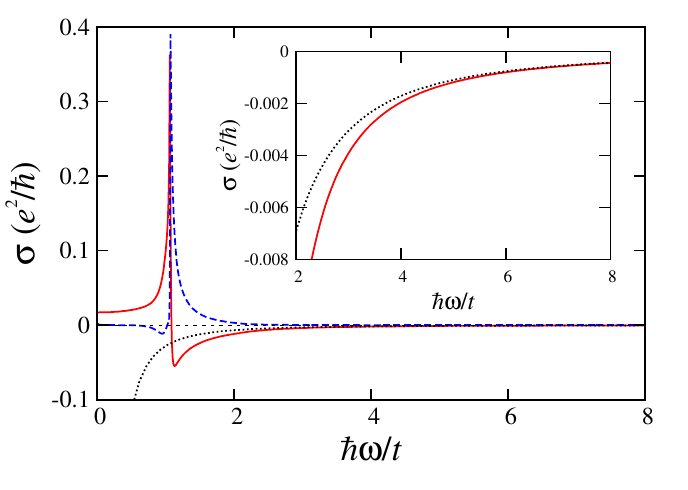}
\caption{(Color online) Real (solid red line) and imaginary (dashed
  blue line) parts of the Hall conductivity calculated using 
  \eq{eq:sigmaH}. The high-frequency approximation from 
  \eq{eq:approxHall} (black dotted line)
  agrees very well with the exact result for $\hbar\omega\gtrsim 4t$.
  The inset compares the approximate and exact results in more detail.  
  We use the same parameters as in~\fig{fig:gappeddispersion}(a) 
  and set the temperature $k_BT=0.05t$.}~\label{fig:Hall} 
\end{figure}

As shown by the Feynman diagrams in~\fig{fig:bubble},
the Hall conductivity can be obtained as the difference 
  \beq
  \sigma_H(\omega)=\frac{i}{2\hbar\omega}\lim_{i\omega_n\to\omega+i0^+}
  \left[\pi_{xy}(i\omega_n)-\pi_{yx}(i\omega_n)\right]
  \eeq
of the current-current correlation functions 
  \beq
  \pi_{ab}(i\omega_n) = -\frac{1}{S}\int^{\hbar\beta}_0 d\tau e^{i\omega_n\tau}
  \left\langle T_\tau J_a(\tau)J_b(0)\right\rangle,
  \eeq
where $\omega_n$ is a bosonic Matsubara frequency and $S$ is the total
  area of the crystal.
Here $J_a$  is the $a$-component of the current operator
\beq
    {\bm J} = e\sum_{\bm k}\Psi^\dagger_{\bm k}{\bm V}_{\bm k}\Psi_{\bm k},
\eeq
where ${\bm V}_{\bm k}$ is the velocity vertex in Nambu notation
\beq
{\bm V}_{\bm k} = {\tau}_0\otimes\left(\begin{array}{cc}0 & {\bm v}_{\bm k}\\
{\bm v}_{\bm k}^\ast & 0 \end{array}\right),
\label{eq:V}
\eeq
and the velocity components are obtained from \eq{eq:epsilon-xy}.
\beq  
{\bm v}_{\bm k}=\frac{1}{\hbar}
\frac{\partial}{\partial\bm k}(\epsilon^x_{\bm k}-i\epsilon^y_{\bm k})
  = -\frac{it}{\hbar} \sum_j \bm R_j e^{i\bm k\cdot\bm R_j}
\label{eq:v}
\eeq
A straightforward evaluation of the Feynman diagrams
in~\fig{fig:bubble} (for the vanishing Semenoff term $\delta_s=0$)
yields the Hall conductivity 
\begin{widetext}
\beq
\sigma_H(\omega) =  \lim_{i\omega_n\to\omega+i0^+}
  \frac{1}{\beta}\sum_{\nu_m}\int\frac{d^2k}{(2\pi)^2}
  \frac{e^2 \hbar^3\mu \, i[{\bm v}_{\bm k}^*\wedge{\bm v}_{\bm k}] \, 
  \Xi_\pm({\bm k}) \, (i\omega_n + 2i\nu_m)^2}
  {(\hbar^2\nu_m^2+E_{{\bm
      k},1}^2)(\hbar^2\nu_m^2+E_{{\bm k},2}^2)(\hbar^2[\omega_n+\nu_m]^2+E_{{\bm
      k},1}^2)(\hbar^2[\omega_n+\nu_m]^2+E_{{\bm k},2}^2)}.
  \label{eq:sigmaH}
\eeq
\end{widetext}
The sign of $\sigma_H(\omega)$ 
correlates with the sign of the chemical potential $\mu$, and the
Hall conductivity vanishes at $\mu=0$ (at the Dirac point) due to
particle-hole symmetry. The real and imaginary parts of the Hall
conductivity calculated from \eq{eq:sigmaH} are shown
in~\fig{fig:Hall}. This expression is consistent with Eq.~(24) 
of~\Ref{KerrUPt3_theory} for the Hall conductivity in UPt$_3$ in
the limiting case where spin-orbit coupling and 
intrasublattice hopping terms are neglected. 
As the point groups of UPt$_3$ and the honeycomb lattice are both $D_{6h}$, such terms are also allowed in our model, but we neglect them for simplicity.

Equation (\ref{eq:sigmaH}) shows some similarity to
a general formula\cite{Tewari2008} for the intrinsic ac Hall conductivity 
in terms of the Berry curvature for a nonsuperconducting two-band system, 
which includes Haldane's model.  However,  that formula is not directly 
applicable to our superconducting case, because the effective 
two-band model derived in Sec.~\ref{sec:massgap} is only suitable
near to the Dirac points.

%%%%%%%%%%%%%%%%%%%%%%%%%%%%%%%%%%%%%%%%%%%%%%%%
%%%%%%%%%%%%%%%%%%%%%%%%%%%%%%%%%%%%%%%%%%%%%%%%

%%%%%%%%%%%%%%%%%%%%%%%%%%%%%%%%%%%%%%%%%%%%%%%%
%%%%%%%%%%%%%%%%%%%%%%%%%%%%%%%%%%%%%%%%%%%%%%%%

From the numerator of \eq{eq:sigmaH}, it is clear that the anomalous
Hall conductivity $\sigma_H(\omega)$ is nonzero only when the
sublattice polarization  $\Xi_\pm(\bm k)$ in \eq{eq:polarization} has common irreducible representations with the product of velocities
$ [{\bm v}_{\bm k}\wedge{\bm v}_{\bm k}^*]$.
Indeed, we find from \eq{eq:v} that
\beq
[{\bm v}_{\bm k}\wedge{\bm v}_{\bm k}^*]
=\frac{2it^2a^2}{\hbar^2} \sum_{j<j'}
 \sin(\varphi_{j'}-\varphi_j) \, \sin[\bm k\cdot(\bm R_j-\bm R_{j'})],
\label{eq:vxv*}
\eeq
where $\varphi_j$ are the geometric angles between the vectors $\bm R_j$ and the $x$ axis, which, in our model, are the same as the phases $\phi_j$ in \eq{eq:gapmatrix}.  Thus, \eq{eq:vxv*} has the same momentum dependence as the TROB in \eq{eq:gapprod} and $\Xi_\pm(\bm k)$ in \eq{eq:polarization}.

The full result~\eq{eq:sigmaH} is rather complicated, but it
simplifies in the high-frequency limit $\hbar\omega \gg t$.  This regime is experimentally relevant, as the polar Kerr effect measurements detailed in~\Ref{KapNJP2009} are performed at infrared frequency $\hbar\omega=0.8$eV which is typically large compared to the hopping integrals in a strongly-correlated material.  As shown in \Ref{Shastry1993}, the Hall conductivity in this limit is given by 
\beq
\sigma_{H}(\omega) \approx
\frac{i}{S\hbar\omega^2}\langle
\left[J_x,J_y\right]\rangle \, . \label{eq:opticalhighw}
\eeq
Taking into account \eq{eq:vxv*}, we find that the commutator 
of the $x$- and $y$-components of the current operator appearing in this expression is directly proportional to the loop-current operator  in \eq{eq:loopcurrent} 
\beq
  \left[J_x,J_y\right] = e^2\sum_{\bm k}
  { [{\bm v}_{\bm k}\wedge{\bm v}_{\bm k}^*]} \,
  \Psi^\dagger_{\bm k}{\tau}_0\otimes{s}_z\Psi_{\bm k} 
  = \frac{\sqrt{3} i e^2t^2a^2}{2\hbar^2}\chi_{lc}\,.
  \label{eq:JxJy}
\eeq
We hence find that the high-frequency Hall conductivity is
proportional to the expectation value of the loop-current operator: 
\begin{align}
 & \sigma_H(\omega) \approx
  -\frac{\sqrt{3}e^2t^2a^2}{2S\hbar^3\omega^2}\langle \chi_{lc}\rangle \label{eq:approxHall} \\
  &  = \frac{2 e^2 }{\hbar\omega^2} \frac{1}{\beta}\sum_{\nu_m}
  \int\frac{d^2k}{(2\pi)^2}
  \frac{\mu i[{\bm v}_{\bm k}^*\wedge{\bm v}_{\bm k}] \, \Xi_\pm({\bm k})
  }{(\hbar^2\nu_m^2 + E_{{\bm k},1}^2)(\hbar^2\nu_m^2 + E_{{\bm k},2}^2)}\,. 
  \nonumber
\end{align}
Equations (\ref{eq:opticalhighw})--(\ref{eq:approxHall})
establish a direct connection between the Hall conductivity and the
loop currents discussed in Sec.~\ref{sec:massgap}. As shown
in~\fig{fig:Hall}, the agreement between Eqs.~(\ref{eq:sigmaH}) and
(\ref{eq:approxHall}) is very good for $\hbar\omega\gtrsim 4t$.
An alternative derivation is presented in Appendix~\ref{eq:diagrams}
in the limit of small $\Delta$, where the Green's functions appearing in 
\fig{fig:bubble} can be expanded to the second order in the pairing
potential. This approach yields \eq{eq:sigmaH-smallD} similar to
\eq{eq:approxHall}, but with the BdG energies $E_{\bm k}$ replaced by
the normal-state energies $\epsilon_{\bm k}$. 

The high-frequency Hall conductivity in \eq{eq:approxHall} is real, but
the polar Kerr effect is primarily sensitive to the imaginary part of
the Hall conductivity when the refraction index is predominantly
real.~\cite{KapNJP2009} Although it is not possible to directly
associate the imaginary part of the Hall conductivity at a given
frequency to the loop currents in the superconductor, an indirect
connection is  provided by the sum rule~\cite{Lange1999}
\beq
\int\limits^{\infty}_{-\infty}\omega
\text{Im}\{\sigma_H(\omega)\}d\omega= - \frac{i\pi}{\hbar S} \langle
\left[J_x,J_y\right] \rangle  =  \frac{\pi\sqrt{3}e^2t^2a^2}{2S\hbar^3}\langle \chi_{lc}\rangle\,.  \notag
\eeq
Again, the right-hand side of this equation is proportional to the
expectation value of the loop-current operator, and we hence conclude that
the existence of the loop-current correlations results in a
nonzero imaginary Hall conductivity.  

It should be noted that, in contrast to nonsuperconductors, the
dc Hall conductivity in superconductors is not directly related to the Chern
number, as discussed in Appendices A and B of \Ref{Lutchyn2009}.  Thus, the
topological phase diagram shown in \fig{fig:Chern} in terms of the
Semenoff term $\delta_s$ is not particularly relevant for the
calculation of the Hall conductivity. A generalization of
\eq{eq:sigmaH} to a nonzero Semenoff term in Appendix~\ref{sec:genexp}
shows that the ac Hall conductivity $\sigma_H(\omega)$ is nonzero for
any value of $\delta_s$. Moreover, in the high-frequency limit,
Eqs.~(\ref{eq:opticalhighw}) and (\ref{eq:JxJy}) are still valid for
$\delta_s\neq0$, so the Hall conductivity remains proportional to the
expectation value of the loop-current operator, which is mainly
sensitive to the pairing potential 
and only weakly dependent upon the Semenoff term.

%%%%%%%%%%%%%%%%%%%%%%%%%%%%%%%%%%%%%%%%%%%%%%%%
\section{Phenomenological treatment} \label{sec:phenomeno}
%%%%%%%%%%%%%%%%%%%%%%%%%%%%%%%%%%%%%%%%%%%%%%%%

In \eq{eq:lc_expectation} we obtained a nonzero expectation value of
the loop-current operator from a microscopic theory of the chiral
$d$-wave state
at the level of the single-particle Green's functions.
The appropriate interactions would, however, lead to true
long-range loop-current order. The interplay of this order with the
superconductivity could then be understood within the framework of a
phenomenological Landau expansion of the  free energy density
\beqarray
f - f_0 &=& \alpha(|\eta_1|^2+|\eta_2|^2) + \beta_1(|\eta_1|^2+|\eta_2|^2)^2
\notag \\
&& +
\beta_2(\eta_1^\ast\eta_2-\eta_1\eta_2^\ast)^2 \notag \\
&& + i\gamma \delta_{lc}(\eta_1^\ast\eta_2-\eta_1\eta_2^\ast) + \kappa \delta_{lc}^2\,,
\eeqarray
where $f_0$ is the normal-state  free energy density.
The first two lines describe the superconductivity, where
$\eta_1$ and $\eta_2$ are the order parameters corresponding to the
two states in the $E_{2g}$ irreducible representation.
The term with $\beta_2>0$
stabilizes the time-reversal symmetry-breaking configuration 
$(\eta_1,\eta_2)\propto(1,\pm i)$ studied  in this paper. 
The coupling to the loop-current order parameter
$\delta_{lc}\propto \langle \chi_{lc}\rangle$
is given by $\gamma$, and $\kappa>0$ implies that this order is
subdominant. Minimization of $f$ with respect to $\delta_{lc}$ shows
that the loop-current order becomes induced in the
time-reversal-breaking superconducting state. 

As already mentioned in Sec.~\ref{sec:TROB-hex}, a more intriguing possibility could be that the loop-current order {\it preempts} the
superconductivity. We speculate that fluctuating
superconducting order may cause the  discrete $\mathbb{Z}_2$ time-reversal
symmetry to be broken  with $\delta_{lc}\neq0$ at a higher
temperature than the continuous $U(1)$ gauge symmetry, which is
rigorously permitted only at zero temperature in two dimensions. 
Similar scenarios were discussed for multiband superconductors in \Ref{Babaev} and for pair-density wave order in the underdoped cuprates in \Ref{cuprate}.

%%%%%%%%%%%%%%%%%%%%%%%%%%%%%%%%%%%%%%%%%%%%%%%
\section{Relevance to superconductivity in twisted bilayer graphene}
\label{sec:TBLG}
%%%%%%%%%%%%%%%%%%%%%%%%%%%%%%%%%%%%%%%%%%%%%%%%%

It has been proposed
theoretically~\cite{Wu2018,Xu2018,Fidrysiak2018,Guo2018,Su2018,Liu2018}
that the superconducting state observed in twisted bilayer
graphene~\cite{Cao2018} (TBLG) realizes chiral $d$-wave pairing.
Given that our analysis concerns hypothetical superconductivity in
monolayer graphene, it is worthwhile to survey theories of TBLG
briefly and explore possible links to our work.  Although most
proposals include more electronic degrees of freedom than our model,
in many cases they show a qualitative resemblance, implying that the
physics discussed in our paper may be applicable.  

Some of the earliest proposals, such as Refs.~\onlinecite{Xu2018,Fidrysiak2018}, assume $SU(4)$ symmetry of a single-particle Hamiltonian, for which the physics we discuss does not apply.  However, $SU(4)$ symmetry-breaking terms may change this conclusion and are currently under consideration. 

Phenomenological models with orbital or sublattice degrees of freedom have been considered in Refs.~\onlinecite{Dodaro2018} and~\onlinecite{Guo2018}, respectively.  Due to the presence of these additional electronic degrees of freedom, the pairing potential may have a nonzero TROB, thus resulting in similar physics to that discussed here.  The model of Ref.~\onlinecite{Guo2018}, based upon a three-site sublattice to simulate the AA and AB regions of the Moir\'{e} pattern of TBLG, resembles our model most closely and, indeed, reduces to it in the limit $t'=\Delta'_j=0$, where the triangular lattice of the AA regions
is neglected. 

Several papers~\cite{Quan2018,Koshino2018,Kang2018,Po2018} proposed a low-energy description of the normal-state electronic structure based on an emergent honeycomb lattice with two additional electronic orbital degrees of freedom at each lattice site.  Reference~\onlinecite{Liu2018} considers such a model with $p_x$ and $p_y$ orbitals on a honeycomb lattice with local electronic interactions and finds that a $d$-wave chiral pairing state emerges.  The paper argues that the mechanism is similar to that for chiral $d$-wave pairing in single-layer graphene at quarter doping, for which our model applies.

Adopting an alternative approach, Refs.~\onlinecite{Su2018}
and~\onlinecite{Wu2018} numerically analyze how a nearest-neighbor
chiral $d$-wave state in each of the two graphene layers is modified
by the Moir\'{e} structure in TBLG.  Interestingly, these papers find
intra-unit-cell supercurrent loops.  Since the intralayer pairing
state is identical to ours, we suggest that these currents may
  be related to the loop current correlations found in our work.

Although a theoretical description of the superconducting state in
TBLG remains unsettled, the presence of multiple electronic degrees of
freedom indicates that a chiral $d$-wave state is likely to have a
nonvanishing TROB.  Consequently, much of the physics discussed in our
paper may be applicable.  An experimental measurement of the polar
Kerr effect in TBLG would be particularly useful to verify whether the
superconducting pairing breaks time-reversal symmetry.

%%%%%%%%%%%%%%%%%%%%%%%%%%%%%%%%%%%%%%%%%%%%%%%%
\section{Conclusions}   \label{sec:conclusions}
%%%%%%%%%%%%%%%%%%%%%%%%%%%%%%%%%%%%%%%%%%%%%%%%

In this paper we have examined the appearance of the polar Kerr effect
in a minimal model of time-reversal symmetry-breaking chiral $d$-wave
superconductivity on the honeycomb lattice. We have demonstrated that
the existence of a gauge-invariant time-reversal-odd bilinear
(TROB) constructed from the pairing potential is an
essential requirement for the polar Kerr effect. In the context of the
honeycomb lattice, the TROB reflects the sublattice polarization  
of the pairing.  The key physical manifestation of the TROB
is the appearance of an emergent nonsuperconducting order in conjunction with
the superconductivity, which we identify as loop currents similar to
those in Haldane's model of a quantum anomalous Hall
insulator.~\cite{Haldane1988} This
is directly evidenced in the energy spectrum, 
where we observe the opening of a topological gap with opposite signs 
at the Dirac points $K$ and $K'$.
The Kubo formula calculation of the intrinsic ac Hall conductivity in
the absence of an external magnetic field shows that it is
directly proportional to the expectation value of the loop current
operator. Thus 
we establish an explicit relation 
connecting the emergent loop-current correlations and both the
real and imaginary parts of the Hall conductivity.

The model considered here is another example of a 
time-reversal symmetry-breaking superconducting state with an
intrinsic Hall conductivity, and generalizes these analyses to an
even-parity pairing state. The first example is Sr$_2$RuO$_4$,
where different pairing in the Ru $d_{xz}$ and 
$d_{yz}$ orbitals implies a polarization in the
$d_{xz}$-$d_{yz}$ orbital space.~\cite{Wysokinski2012,Taylor2012,GradhandSr2RuO4,RobbinsSr2RuO4,KomendovaPRL2017} 
More recently, a theoretical  treatment of the Kerr effect in UPt$_3$ has identified the time-reversal-odd sublattice dependence of the 
pairing potential permitted by the nonsymmorphic
symmetry as an essential ingredient.~\cite{KerrUPt3_theory,Yanase2016}
The similarity of these models to the simpler
case considered  in our paper suggests the intriguing possibility that the
Hall conductivities in these systems can be also understood in terms of
loop-current correlations induced by a TROB. Although we have
considered only two internal degrees of freedom here, loop currents
can also arise in materials with more complicated 
unit cells,~\cite{Ghosh2018} including twisted bilayer graphene. 
The observation of the polar Kerr effect in many unconventional
superconductors therefore suggests that pairing states 
supporting nonzero TROBs may be realized in a broad range of materials.

%%%%%%%%%%%%%%%%%%%%%%%%%%%%%%%%%%%%%%%%%%%%%%%%
\begin{acknowledgments}
The authors acknowledge useful discussions with Fengcheng Wu,
 Carsten Timm and Henri Menke. We thank  the  hospitality  of  the  summer 2016 program  
``Multi-Component  and Strongly-Correlated Superconductors'' at Nordita,
Stockholm, where this work was initiated.  DSLA acknowledges support of
ERC project DM-321031.  DFA acknowledges support from the NSF via DMREF-1335215.
\end{acknowledgments}
%%%%%%%%%%%%%%%%%%%%%%%%%%%%%%%%%%%%%%%%%%%%%%%%

\appendix

%%%%%%%%%%%%%%%%%%%%%%%%%%%%%%%%%%%%%%%%%%%%%%%%
\section{Chiral $p$-wave state}
\label{sec:chiralp}
%%%%%%%%%%%%%%%%%%%%%%%%%%%%%%%%%%%%%%%%%%%%%%%%

\begin{figure}[t!]
(a)\includegraphics[width=0.45\columnwidth]{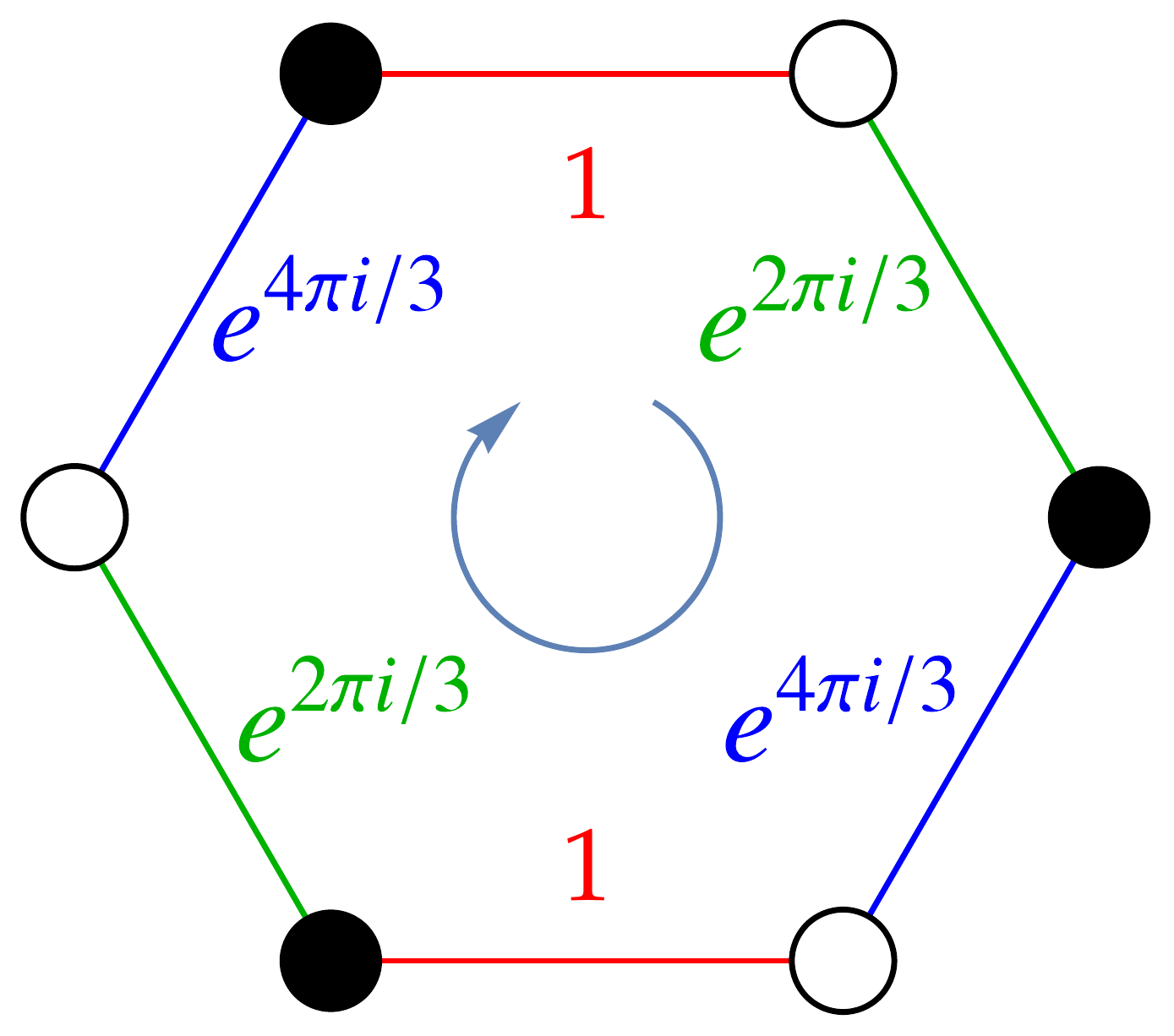}(b)\includegraphics[width=0.45\columnwidth]{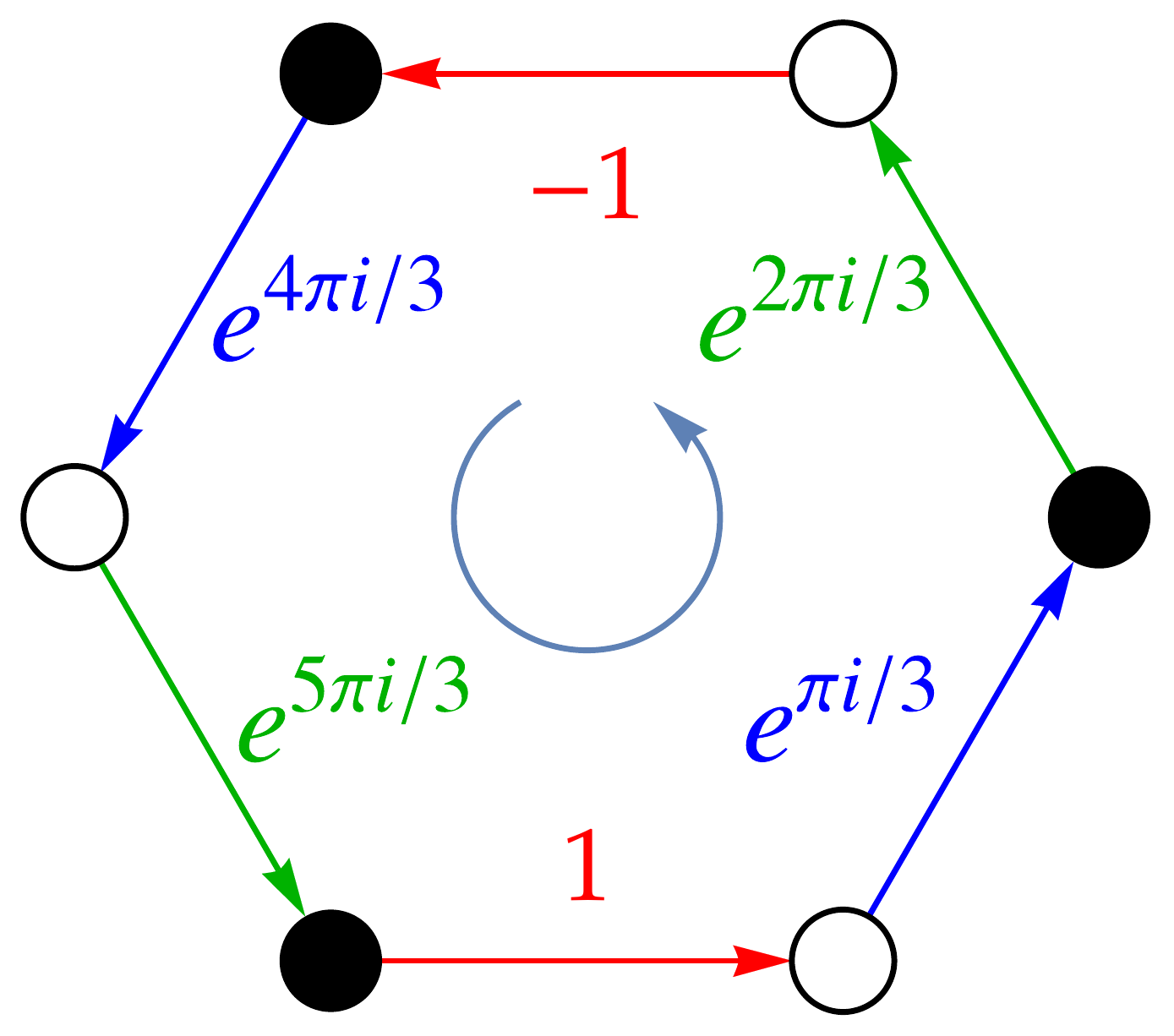}
\caption{(Color online) Phases of the nearest-neighbor pairing
  potentials on the bonds of a hexagonal 
  plaquette for the (a) chiral $d$-wave state and (b) the 
  chiral $p$-wave state. In the former the phase winds by $4\pi$
  around the plaquette, whereas in the latter in winds by $2\pi$, and 
  the direction of winding is indicated by the curved arrow. In the
  spin-triplet $p$-wave state, the straight arrow points to the
  location of the spin-$\downarrow$ electron 
  in the Cooper pair for the given phase. 
  Placing the spin-$\downarrow$
  electron at the other end of the arrow gives an overall sign change of the
  pairing potential. }~\label{fig:chiral} 
\end{figure}

Other chiral pairing states on the honeycomb lattice also display a
polarization in their sublattice degrees of freedom due to a non-zero
TROB. For example, chiral $p$-wave triplet  superconductivity with
nearest-neighbor pairing has been considered by several  
authors.~\cite{Xu2016,Faye2015,QiFuSunGu2017,Ying2017} 
Here we examine the case where the vector $\bm d$ of the triplet
pairing is oriented along the $z$ axis, which is also the spin
quantization axis.  In this representation, the triplet pairing takes
place between the opposite spins and is described by 
the BdG Hamiltonian in \eq{eq:HBdG} with the pairing potential
\beq 
  {\Delta}_{\pm}^{p}({\bm k}) =  
  \Delta \sum\limits_{j=1}^3 e^{\pm i\phi_j} \left(\begin{array}{cc} 0 &
  e^{i{\bm k}\cdot{\bm R}_j} \\
  - e^{-i{\bm k}\cdot{\bm R}_j} & 0 \end{array}\right)\,.  \label{eq:gapmatrix-p}
\eeq
Note that one of the off-diagonal terms has opposite sign compared
with~\eq{eq:gapmatrix} for the pairing potential of the chiral $d$-wave
state, but the phases along each bond $\phi_j=(j-1)2\pi/3$ are the
same. As shown in~\fig{fig:chiral}, the 
phase of the pairing on each bond winds by
$2\pi$ as one moves around a hexagonal plaquette, in contrast to the
chiral $d$-wave state where the phase winds by $4\pi$. 

%%%%%%%%%%%%%%%%%%%%%%%%%%%%%%%%%%%%%%%%%%%%%%%%
%%%%%%%%%%%%%%%%%%%%%%%%%%%%%%%%%%%%%%%%%%%%%%%%

The pairing potential in \eq{eq:gapmatrix-p}
can be decomposed into basis states of the
irreducible representation $E_{1u}$
\beq
{\Delta}^{p}_{\pm}({\bm k}) = {\Delta}^{p}_x({\bm k}) \pm i{\Delta}^{p}_y({\bm k}),
\eeq
where
\beqarray
{\Delta}^{p}_x({\bm k}) & = &
i\Delta\left\{\left[\cos(k_xa) -
  \cos(\tfrac{1}{2}k_xa)\cos(\tfrac{\sqrt{3}}{2}k_ya)\right]{s}_y\right. \notag
\\
&& \left.
+ \left[\sin(k_xa) +
  \sin(\tfrac{1}{2}k_xa)\cos(\tfrac{\sqrt{3}}{2}k_ya)\right]{s}_x\right\},
\notag \\
{\Delta}^{p}_y({\bm k}) & = &
i\sqrt{3}\Delta\left\{\sin(\tfrac{1}{2}k_xa)\sin(\tfrac{\sqrt{3}}{2}k_ya){s}_y
\right.\notag \\
&& \left.
+\cos(\tfrac{1}{2}k_xa)\sin(\tfrac{\sqrt{3}}{2}k_ya)
{s}_x\right\}. \notag
\eeqarray
Projected onto the states at the Fermi surface, the basis functions
${\Delta}^{p}_x({\bm k})$ and ${\Delta}^{p}_y({\bm k})$
appear as $p_x$-wave and $p_y$-wave triplet states, respectively.
Like the basis functions for $E_{2g}$ discussed in Sec,~\ref{sec:model},
these states contain the matrices ${s}_x$ and ${s}_y$, but here with
odd- and even-parity coefficients, respectively.  This ensures that
the  pairing potentials are odd under inversion, i.e.\
${\cal  
  I}^\dagger \Delta^p_{x}({\bm k}) {\cal I} = -\Delta^p_{x}(-{\bm
  k})$.

The $E_{1u}$ basis functions can be obtained from the basis
functions for $E_{2g}$ by multiplying them with ${s}_z$. 
This follows from the direct product
rules for the point group $D_{6h}$, since $s_z$ belongs to the
irreducible representation $B_{1u}$ and $E_{1u}=B_{1u}\times E_{2g}$.
Thus, the TROB of the chiral $p$-wave state is the same as for the chiral $d$-wave case in \eq{eq:gapprod}:  
\begin{align}
  \lefteqn{\mbox{TROB}}\notag \\
  & = \mp 4\sqrt{3} |\Delta|^2
\sin(\tfrac{\sqrt{3}}{2}k_ya)\left[\cos(\tfrac{3}{2}k_xa)-\cos(\tfrac{\sqrt{3}}{2}k_ya)\right]{s}_z\,. \notag
\end{align}
The physics arising from the existence of the TROB in the chiral
$p$-wave state is thus
essentially the same as for the chiral $d$-wave state discussed
in the main part of the paper.

We finally note that, in the presence of a Semenoff term, the reduced
symmetry of the lattice  due to lack of inversion implies that both the $p$-wave and $d$-wave pairing potentials are basis states of the same irreducible
representation $E^\prime$ of the point group $D_{3h}$. As such, a
chiral state can generally involve a mixture of the
two.~\cite{BlackSchaeffer2015,QiFuSunGu2017}

%%%%%%%%%%%%%%%%%%%%%%%%%%%%%%%%%%%%%%%%%%%%%%%%
\section{Time-reversal odd bilinear in more general models}
\label{sec:TROBgeneral}
%%%%%%%%%%%%%%%%%%%%%%%%%%%%%%%%%%%%%%%%%%%%%%%%

Here we show that \eq{eq:TROB} for the TROB applies quite generally, including Hamiltonians with spin-orbit coupling and more electronic degrees of freedom,
and without inversion symmetry. 

The pairing term in \eq{eq:HBdG} couples opposite spins, but a more general BdG Hamiltonian can be written in terms of a $2m$-component operator
$c_{\bm k}$ encoding $m$ orbital or sublattice degrees of freedom (assumed to be time-reversal invariant) and both spin orientations:
\begin{equation} \label{eq:Hamiltonian}
H = \frac{1}{2} \sum_{\bm k}\Psi^\dagger_{\bm k}{\cal H}_{\bm k}\Psi_{\bm k}\,.
\end{equation}
Here $\Psi_{\bm k}=(c_{\bm k},c^\dagger_{-{\bm k}})^T$ is a Nambu spinor, and $T$ denotes transposition.  We use $\hat\Delta(\bm k)$ with a ``hat'' to denote pairing potential in this basis:
\begin{equation}
{\cal H}_{\bm k} = \left(\begin{array}{cc}
  H_0(\bm k) & \hat\Delta(\bm k)\\
  \hat\Delta^\dagger(\bm k) & -H_0^T(-\bm k)
  \end{array}\right) . \label{eq:BdGH}
\end{equation}
The pairing potential obeys $\hat\Delta^T(\bm k)=-\hat\Delta(-\bm k)$
due to the fermion exchange symmetry.~\cite{SPOT}

The time-reversal operation is implemented as $\Theta = {\cal K}U$, where $\cal K$ is complex conjugation, and the unitary part~\cite{Agterberg2017} is
\beq \label{eq:U}
U=i\sigma_y \otimes s_0 = 
\left(\begin{array}{rr} 0 & 1 \\ -1 & 0\end{array}\right) \otimes s_0\,.
\eeq
Here $\sigma_y$ is the spin Pauli matrix, and $s_0$ is an $m$-dimensional identity matrix operating on the $m$ orbital or sublattice electronic degrees of freedom.
The matrix $U$ is real and satisfies $U^2=-1$ and $U^\dagger=U^T=U^{-1}=-U$.  Because it is real, the creation and annihilation operators transform upon time reversal in the same way, as discussed in Sec.~\ref{sec:TROB-hex}:
\beqarray
c_{\bm k} &\;\to\;& \Theta c_{\bm k} \Theta^{-1}=Uc_{-\bm k} \notag \\
c^\dagger_{\bm k} &\;\to\;& \Theta c^\dagger_{\bm k} \Theta^{-1}=Uc^\dagger_{-\bm k} \notag \\
\Psi_{\bm k} &\;\to\;& \Theta \Psi_{\bm k} \Theta^{-1}=[\tau_0\otimes U]\,\Psi_{-\bm k}\,, \notag
\eeqarray
where $\tau_0$ is a $2\times 2$ identity matrix in the Nambu space.
The matrix elements in \eq{eq:BdGH} become complex-conjugated upon time reversal:
${\cal H}_{\bm k}\to \Theta{\cal H}_{\bm k}\Theta^{-1}={\cal H}_{\bm k}^*$.  Combining these transformations, we obtain the time reversal of the BdG Hamiltonian (\ref{eq:Hamiltonian})
\beqarray
H &\;\to\;& \frac{1}{2} \sum_{\bm k} \Theta\Psi^\dagger_{\bm k}\Theta^{-1} 
\Theta{\cal H}_{\bm k}\Theta^{-1} \Theta\Psi_{\bm k}\Theta^{-1} \notag \\
&=& \frac{1}{2} \sum_{\bm k}  \Psi_{-\bm k}^\dagger\, [\tau_0\otimes U^T]\,
{\cal H}_{\bm k}^*\, [\tau_0\otimes U]\, \Psi_{-\bm k}\,. \notag
\eeqarray
Changing $\bm k\to -\bm k$ in the sum, we arrive to a BdG Hamiltonian of the same form as in \eq{eq:Hamiltonian} but with the transformed matrix elements in \eq{eq:BdGH}
\begin{equation} \label{eq:time-reversal} 
{\cal H}_{\bm k} \;\to\; [\tau_0\otimes U^T]\, {\cal H}^*_{-\bm k}\, [\tau_0\otimes U].
\end{equation}
The time-reversal rule in \eq{eq:time-reversal} is similar to that for a non-superconducting Hamiltonian,~\cite{Bernevig} because $U$ is real.

According to \eq{eq:time-reversal}, the pairing potential transforms as 
\begin{equation} 
\hat\Delta(\bm k) \;\to\; U^T\hat\Delta^*(-\bm k)U
=-U^T\hat\Delta^\dagger(\bm k)U, \label{eq:TRhatD} 
\end{equation}
where we used the fermion exchange relation.  It is convenient to define
$\Delta({\bm k})$ without the ``hat'' as 
\begin{equation}  \label{eq:hat}
\Delta({\bm k})=\hat\Delta({\bm k})U,
\end{equation}
which corresponds more closely to the pairing potential introduced in Sec.~\ref{sec:model}.  Combining \eq{eq:TRhatD} with \eq{eq:hat} and using $U^2=-1$, we reproduce \eq{D->D+}
\beq  \label{D->D+app}
\Delta(\bm k)\to\Delta^\dagger(\bm k).
\eeq 

Finally, we obtain the TROB as the difference between 
$\hat\Delta({\bm k})\hat\Delta^\dagger({\bm k})$ 
and its time-reversed counterpart:
\beqarray  \notag   
 \mbox{TROB} &=& \hat\Delta({\bm k})\hat\Delta^\dagger({\bm k})
 -\Theta\hat\Delta({\bm k})\Theta^{-1}\Theta\hat\Delta^\dagger({\bm k})\Theta^{-1} \\
 &=& \hat\Delta({\bm k})\hat\Delta^\dagger({\bm k})
 -U^T\hat\Delta^{\dagger}({\bm k})\hat\Delta({\bm k})U \notag \\
 &=& \Delta(\bm k)\Delta^\dagger(\bm k) - \Delta^\dagger(\bm k)\Delta(\bm k),
 \label{eq:troddgapprod}
\eeqarray
where we used Eqs.\ (\ref{eq:TRhatD}) and (\ref{eq:hat}).  Equation (\ref{eq:troddgapprod}) reproduces \eq{eq:TROB} for the TROB,
which is the result we wanted to show in general form.

As discussed in Sec.~\ref{sec:TROB-hex}, the superconducting fitness restricts opportunities for a nonzero TROB.  The general condition for superconducting fitness~\cite{Ramires2016,Ramires2018} is 
\beq \label{eq:BestFit} 
H_0(\bm k)\hat\Delta(\bm k)=\hat\Delta(\bm k)H_0^*(-\bm k).
\eeq
If the normal-state Hamiltonian $H_0(\bm k)$ is invariant under 
time reversal in \eq{eq:time-reversal}, then, using \eq{eq:hat}, 
\eq{eq:BestFit} reduces to the commutator in \eq{eq:BestFitCommut}
\beq \label{eq:BestFit-again} 
[H_{0}({\bm k}),\Delta({\bm k})]=0.
\eeq  

Below we illustrate these general relations by simple examples.  For a single-band superconductor, where the only internal degree of freedom is spin, the pairing potential can be written in terms of the spin Pauli matrices 
\begin{equation}  \label{eq:single-band}
\Delta({\bm k})=\Delta^{(s)}_{\bm k}
+\Delta^{(t)}_{\bm k} ({\bm d}_{\bm k}\cdot \bm\sigma).
\end{equation}
Here $\Delta^{(s)}_{\bm k}$ and $\Delta^{(t)}_{\bm k}$ represent singlet and triplet pairing, and $\bm d_{\bm k}$ is a unit vector.  Evaluation of
\eq{eq:troddgapprod} for this pairing potential gives 
\begin{equation} \label{eq:TROBsingle-band}
\mbox{TROB} = 2i\, |\Delta^{(t)}_{\bm k}|^2\, 
(\bm d_{\bm k}\times \bm d^*_{\bm k}) \cdot \bm\sigma.
\end{equation}
Clearly, $\mbox{TROB}\neq0$ only when the vector 
$\bm d^*_{\bm k}$ is not parallel to $\bm d_{\bm k}$.
The pairing with $\bm d_{\bm k}\times \bm d^*_{\bm k}\neq0$ is known in the literature~\cite{SigUeda1991} as nonunitary pairing, because the product $\Delta(\bm k)\Delta^\dagger(\bm k)$ is not proportional to the unit matrix.  Obviously, $\mbox{TROB}$ in \eq{eq:troddgapprod} vanishes for a unitary pairing, so nonunitarity is a necessary, but generally not sufficient condition for 
$\mbox{TROB}\neq0$.

A similar construction can be obtained for spin-singlet pairing in a two-band model, where the pairing potential is expanded in the Pauli matrices $s_\lambda$ for sublattice space:
\beq  \label{eq:Dk}
\Delta({\bm k}) =\Delta^{(0)}_{\bm k}s_0+\bm\Delta_{\bm k}\cdot{\bm s}.
\eeq
The honeycomb lattice model described in Sec.~\ref{sec:model} is a special case  where $\Delta^{(0)}=0$ and the vector $\bm\Delta_{\bm k}$ has only two components.
An evaluation of \eq{eq:troddgapprod} for \eq{eq:Dk} gives a formula similar to \eq{eq:TROBsingle-band}
\beq  \label{eq:TROB22}
  \mbox{TROB} = [\Delta({\bm k}),\Delta^\dagger({\bm k})]
  = 2i (\bm\Delta_{\bm k}\times\bm\Delta^*_{\bm k})\cdot{\bm s}.
\eeq
For $\mbox{TROB}\neq0$, the pairing vector $\bm\Delta_{\bm k}$ must be not parallel to its complex conjugate.

To evaluate the fitness condition for this model, we take the normal-state Hamiltonian to be spin-independent and write it as
\beq  \label{eq:h0k}
H_0({\bm k}) = h^{(0)}_{\bm k}s_0+{\bm h}_{\bm k}\cdot{\bm s},
\eeq 
where $h^{(0)}_{\bm k}$ and ${\bm h}_{\bm k}$ are necessarily real, because $H_0$ is Hermitian.  Then the fitness condition (\ref{eq:BestFit-again}) is
\beq  \label{eq:Fit22}
  [H_0({\bm k}),\Delta({\bm k})]
  = 2i [{\bm h}_{\bm k}\times{\bm\Delta}_{\bm k}]\cdot{\bm s}.
\eeq
Perfect fitness is achieved when the pairing vector
${\bm\Delta}_{\bm k}$ is parallel to the real vector ${\bm h}_{\bm k}$, 
which makes TROB vanish in \eq{eq:TROB22}, so perfect fitness is incompatible with $\mbox{TROB}\neq0$.

%%%%%%%%%%%%%%%%%%%%%%%%%%%%%%%%%%%%%%%%%%%%%%%%
\section{Effect of Semenoff term}
\label{sec:genexp} 
%%%%%%%%%%%%%%%%%%%%%%%%%%%%%%%%%%%%%%%%%%%%%%%%

For simplicity, the main text gives~\eq{eq:lc_expectation} for the
loop-current operator expectation value and~\eq{eq:sigmaH} for the Hall
conductivity only in the absence of the Semenoff term, 
  i.e.\ for $\delta_s=0$. Here we present the general expressions for 
$\delta_s\neq0$, which may be useful for applications to transition
  metal dichalcogenides,~\cite{TiSe2,MoS2,dichalcogenides} where the
  $A$ and $B$ sites are strongly inequivalent. 

In the presence of the Semenoff term, the expectation value of the
loop-current operator is given by 
\beqarray
\frac{\langle \chi_{lc} \rangle}{N} &=& -\frac{1}{N}\sum_{{\bm
    k}}\sin(\tfrac{\sqrt{3}}{2}k_ya)\left[\cos(\tfrac{3}{2}k_xa) -
  \cos(\tfrac{\sqrt{3}}{2}k_ya)\right] \notag \\
&&\times \frac{1}{\beta}\sum_{\nu_m}
\frac{8\mu\,\left[4\delta_si\hbar \nu_m+\text{Tr}\{{\Delta}^\dagger({\bm
    k}){s}_z{\Delta}({\bm k})\}\right]}
    {\hbar^4\nu_m^4 + c_2\hbar^2\nu_m^2 + c_1 i\hbar\nu_m + c_0}\,. 
\label{eq:gen_ls_expectation}
\eeqarray
where the coefficients of the quartic polynomial in the fermionic
frequency $\nu_m$ in the denominator are
\beqarray 
c_2 & = & 2(\delta_s^2+\epsilon_x^2+\epsilon_y^2+\mu^2) +
\text{Tr}\{\Delta^\dagger({\bm k})\Delta({\bm k})\} \notag \\
c_1 & = & -2\delta_s\text{Tr}\{{\Delta}^\dagger({\bm
  k}){s}_z{\Delta}({\bm k})\} \notag \\
c_0 & = & (\delta_s^2+\epsilon_x^2+\epsilon_y^2-\mu^2 - \tfrac{1}{2}\text{Tr}\{{\Delta}^\dagger({\bm
  k}){\Delta}({\bm k})\})^2 \notag \\
&& -  \tfrac{1}{4}|\text{Tr}\{{\Delta}^\dagger({\bm
  k}){s}_z{\Delta}({\bm k})\}|^2 + |\text{Tr}\{H_0({\bm
  k}){\Delta}({\bm k})\}|^2\,. \notag
\eeqarray
The numerator in~\eq{eq:gen_ls_expectation} is no longer directly 
proportional to the sublattice polarization
$\text{Tr}\{{\Delta}^\dagger({\bm k}){s}_z{\Delta}({\bm k})\}$, but
now also contains a term proportional to the Semenoff term
$\delta_s$. Nevertheless, the contribution from this additional term,
which is also proportional to fermionic frequency $\nu_m$, is only 
nonzero if the coefficient $c_1$ of the linear term in the denominator
is also nonzero.  As this is only the case if the pairing potential
has a time-reversal symmetry-breaking sublattice polarization,
the key role of the non-zero TROB in producing the loop 
current correlations is robust to the presence of the Semenoff
term.  

The Hall conductivity in the presence of the Semenoff term is given by
\begin{widetext}
\beqarray
\sigma_H(\omega) & = & 
   \lim_{i\omega_n\to\omega+i0^+} \frac{1}{\beta}
  \sum_{\nu_m} \int\frac{d^2k}{(2\pi)^2} i e^2\mu\hbar^2[{\bm v}_{\bm k}\wedge{\bm v}_{\bm k}^\ast]
    \, (i\omega_n + 2i\nu_m)
  \label{eq:gen_sigmaH} \\
  &&\times \frac{\left[4\delta_s(\delta_s^2+\epsilon_x^2+\epsilon_y^2-\mu^2 - \tfrac{1}{2}\text{Tr}\{{\Delta}^\dagger({\bm
  k}){\Delta}({\bm k})\}-i\hbar^2\nu_m[i\nu_m+i\omega_n])
  - \hbar(i\omega_n+2i\nu_m)\text{Tr}\{{\Delta}^\dagger({\bm
    k}){s}_z{\Delta}({\bm k})\}\right]}
    {(\hbar^4\nu_m^4 + c_2\hbar^2\nu_m^2 + c_1i\hbar\nu_m + c_0)
    (\hbar^4[\omega_n+\nu_m]^4 + c_2\hbar^2[\omega_n+\nu_m]^2 + c_1\hbar[i\omega_n + i\nu_m] + c_0)}
    \,. \notag 
\eeqarray
\end{widetext}
Similarly to~\eq{eq:gen_ls_expectation}, a nonzero Semenoff term again
results in a new term proportional to $\delta_s$ in the numerator. The
coefficient of $\delta_s$ in the numerator of~\eq{eq:gen_sigmaH} has
the full symmetry of the lattice, whereas the prefactor 
$ [{\bm v}_{\bm k}\wedge{\bm v}_{\bm k}^\ast]$ belongs to the
irreducible representation $A_2^\prime$ of the point group
$D_{3h}$. The contribution from this new term will thus be vanishing,
unless the denominator also contains a term in the irreducible
representation $A_2^\prime$. Such a term is only present if the linear
coefficient $c_1$ of the polynomials in the denominator is nonzero,
which requires a sublattice polarization of the pairing.  Thus, the 
nonzero Hall conductivity remains a signature of a finite TROB
in the presence of the Semenoff term. 

%%%%%%%%%%%%%%%%%%%%%%%%%%%%%%%%%%%%%%%%%%%%%%%%
\section{\boldmath
High-frequency small-$\Delta$ limit of the Hall conductivity}
\label{eq:diagrams} 
%%%%%%%%%%%%%%%%%%%%%%%%%%%%%%%%%%%%%%%%%%%%%%%%

\begin{figure*}
    \includegraphics[width=2\columnwidth]{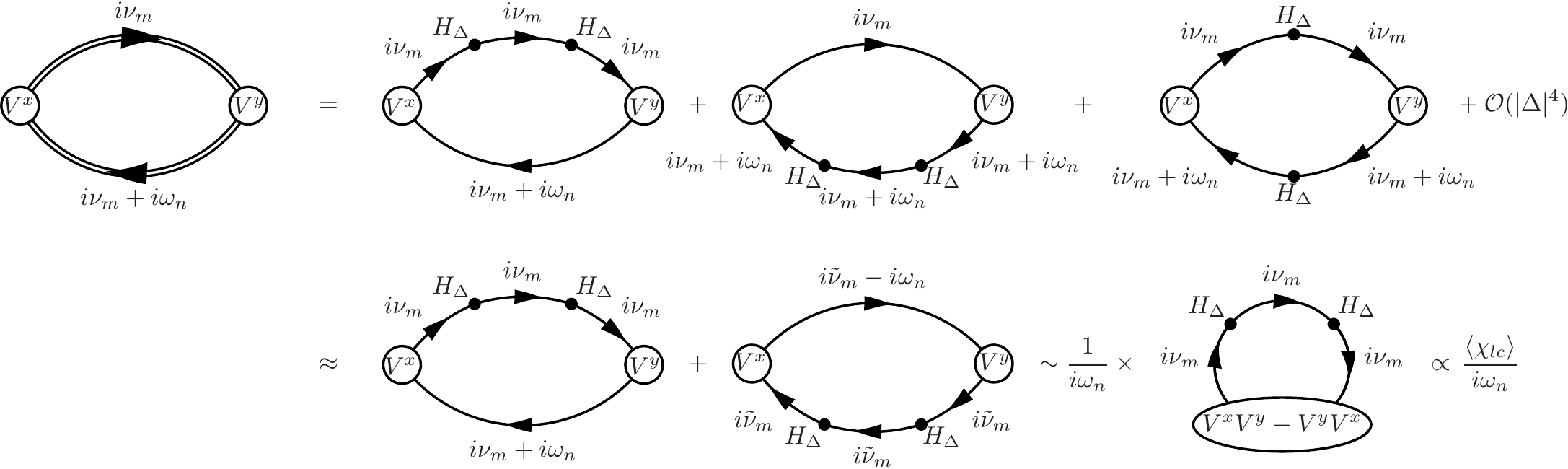}
\caption{Diagrammatic derivation of the high-frequency 
    small-$\Delta$ limit. The current-current correlator $\pi_{xy}(i\omega_n)$
  is expanded in powers of the pairing Hamiltonian ${H}_\Delta$, 
    which is treated as a perturbation. The leading-order
  contribution in the high-frequency limit comes from the first two
  terms on the right hand side, as shown in the second line. Note 
  the redefinition of the internal frequency 
  $i\tilde{\nu}_m = i\nu_m + i\omega_n$ in the second term.
  After the Green's functions containing the external frequency are
  factored out, the result is expressed in terms of the expectation
  value of the loop-current operator $\chi_{lc}$. The double line 
  represents the full Green's function ${G}$, and the single line 
  denotes the Green's function ${G}_0$ of the
  normal system.}\label{fig:bubbleapprox}
\end{figure*}

The high-frequency limit of the Hall conductivity was derived
in~\Ref{Shastry1993} from the general form of the
current-current correlation function. Here we present an alternative
derivation based upon approximation of the Green's functions in the
Feynman diagrams shown in~\fig{fig:bubble}. Specifically, in the high-frequency
limit $|\omega| \gg |\Delta|$, the Hall conductivity~\eq{eq:sigmaH}
should only weakly depend upon the modification of energy spectrum in the
superconducting state. We thus expect that a perturbative expansion in
the pairing Hamiltonian will quickly converge.  
To achieve this, we first note that the full Green's function
${G}$ is related to the Green's 
function of the normal system ${G}_0$ by the Dyson's equation
\beq
{G} = {G}_0 + {G}_0{H}_\Delta{G},
\eeq
where ${H}_\Delta$ is the pairing part of the
BdG Hamiltonian~\eq{eq:HBdG}. Expanding this to second order in
${H}_\Delta$, we approximate ${G}$ by
\beq
{G} \approx {G}_0 + {G_0}{H}_\Delta{ G}_0 +
{G_0}{H}_\Delta{G_0}{H}_\Delta{ G}_0 +
      {\cal O}(|\Delta|^3)\,.  \label{eq:G123}
\eeq
Note that the normal part of the Green's function in \eq{eq:G123}
reproduces \eq{eq:G-1}.  Using the approximate 
\eq{eq:G123} to replace the full Green's function in the
current-current correlator $\pi_{xy}(i\omega_n)$, we obtain the
expansion shown in~\fig{fig:bubbleapprox}. Performing the analytic
continuation $i\omega_n\to\omega + i0^{+}$, the first diagram on the
right hand side is $\sim 1/\omega$
in the high-frequency limit, as the external frequency passes through a
single normal-state Green's function. The next diagram is also 
$\sim 1/\omega$, since a redefinition of the internal frequency (see
  second line) also allows
the external frequency $i\omega_n$ to pass through a single
normal-state Green's function. In contrast, the external frequency in
the third diagram must necessarily pass through two Green's functions,
and this diagram can be shown to be at least $\sim 1/\omega^2$
in the high-frequency limit.

%%%%%%%%%%%%%%%%%%%%%%%%%%%%%%%%%%%%%%%%%%%%%%%%
%%%%%%%%%%%%%%%%%%%%%%%%%%%%%%%%%%%%%%%%%%%%%%%%

Keeping only the first two diagrams, therefore, we approximate
$\pi_{xy}(i\omega_n)$ as shown in the second line
of~\fig{fig:bubbleapprox}. We observe that, in performing the Matsubara
summation over the internal frequency, the residue of the poles of the
Green's function containing the external frequency will be at least
$\sim 1/\omega^3$, whereas the residue of the poles of the other
Green's functions will be 
$\sim 1/\omega$. Since the $\sim 1/\omega$
contribution only arises from the  
unit matrix (i.e.\  ${\tau}_0\otimes{s}_0$) component of the
Green's function containing the external frequency, we make the
approximation 
${G}_0({\bm k}, i\nu_m \pm i\omega_n) \approx (\pm i\omega_n)^{-1}{\tau}_0\otimes{s}_0$
and hence factor the external frequency out of the 
Matsubara sum. This yields the diagram 
involving the commutator of the velocity vertices and the second-order
Green's function correction
${G_0}{H}_\Delta{G_0}{H}_\Delta{ G}_0$.
This product is proportional to the expectation value of the
dimensionless loop-current
operator~\eq{eq:lc_expectation} expanded to lowest order in the
pairing potential. Evaluating this diagram, we obtain the Hall conductivity   
\beqarray
  && \pi_{xy}(i\omega_n) \notag \\
  && \approx 
    \frac{e^2}{\hbar i\omega_n}\frac{1}{S\beta}\sum_{{\bm k},\nu_m}
  \text{Tr}\{(V^x_{\bm k}V^y_{\bm k}-V^y_{\bm k}V^x_{\bm k}) G_0 H_\Delta
  G_0 H_\Delta G_0 \} \notag \\
  && = \frac{e^2}{\hbar i\omega_n} \frac{1}{\beta}\sum_{\nu_m}
  \int\frac{d^2k}{(2\pi)^2}
  \frac{ \mu[\bm v_{\bm k}^*\wedge\bm v_{\bm k}] \, \Xi_\pm(\bm k)}
  {(\hbar^2\nu_m^2 + \epsilon_{\bm k,1}^2)
  (\hbar^2 \nu_m^2 + \epsilon_{\bm k,2}^2)}, \notag
\eeqarray 
where $\epsilon_{{\bm k},1(2)} 
= +(-) \sqrt{(\epsilon^x_{\bm k})^2+(\epsilon^y_{\bm k})^2}-\mu$ 
are the dispersions in the normal state.  A similar analysis yields
$\pi_{yx}(i\omega_n)=-\pi_{xy}(i\omega_n)$.  We hence obtain the Hall conductivity
  \beq \label{eq:sigmaH-smallD}
  \sigma_H(\omega) = \frac{2e^2}{\hbar\omega^2} \frac{1}{\beta}\sum_{\nu_m}
  \int\frac{d^2k}{(2\pi)^2}
  \frac{i\mu[\bm v_{\bm k}^*\wedge\bm v_{\bm k}] \, \Xi_\pm(\bm k)}
  {(\hbar^2\nu_m^2 + \epsilon_{\bm k,1}^2)
  (\hbar^2\nu_m^2 + \epsilon_{\bm k,2}^2)}\,. \notag
  \eeq
We recognize this as the lowest-order term in the expansion of~\eq{eq:approxHall} in powers of the gap magnitude.

%%%%%%%%%%%%%%%%%%%%%%%%%%%%%%%%%%%%%%%%%%%%%%%%

%%%%%%%%%%%%%%%%%%%%%%%%%%%%%%%%%%%%%%%%%%%%%%%%
%%%%%%%%%%%%%%%%%%%%%%%%%%%%%%%%%%%%%%%%%%%%%%%%

\begin{thebibliography}{100}
\bibitem{bilinear} More complicated bilinear products of $\Delta({\bm k})$ and $\Delta^\dagger({\bm k})$ may also involve $H_0(\bm k)$, 
but we focus on the simplest case of $\Delta(\bm k)\Delta(\bm k)^\dagger$
in our paper.
\end{thebibliography}
\end{document}